%% file: paper.tex
\documentclass[]{gensi}
\usepackage{tabularx}
\usepackage[toc,page,header]{appendix}

\usepackage{minitoc}
\usepackage{cleveref} 
\usepackage{microtype}
\usepackage{subcaption}
\usepackage{amsmath} 
\usepackage{amssymb} 
\usepackage{amsfonts}       
\usepackage{pgfplots}
\usepackage{pgfplotstable}
\usepackage{xcolor}
\usepackage{CJKutf8}
\usetikzlibrary{patterns}
\usepackage{multirow}
\usepackage{setspace}
\usepackage{xcolor}
\usepackage{hyperref}
\tcbuselibrary{theorems}
\usepackage{float}
\usepackage{caption}
\usepackage{wrapfig}
\usepackage{xspace}
\usepackage{graphicx}      
\usepackage{subcaption}    
\usepackage{seqsplit}      

\usepackage{siunitx}
\usepackage{soul}          
\usepackage{array}

\usepackage{tcolorbox}
\newtcolorbox{DefinitionBox}{
  colback=blue!5,
  colframe=blue!80,
  boxrule=0.5pt,
  arc=2pt,
  left=2pt,
  right=2pt,
  top=2pt,
  bottom=2pt,
}

\newtcolorbox{CorollaryBox}{
  colback=gray!5,
  colframe=gray!80,
  boxrule=0.5pt,
  arc=2pt,
  left=2pt,
  right=2pt,
  top=2pt,
  bottom=2pt,
}

\usepackage{fix-cm}

\input{macro}

\input{math_commands}

\newcommand{\method}{\textsc{AMix-2}\xspace}

\newcommand{\methodt}{\textsc{AMix-2}}

\newcommand{\bench}{\textsc{ProteinArena}\xspace}

\definecolor{myclr}{RGB}{201, 236, 255}
\newcommand{\cc}{\cellcolor{myclr}} 

\title{
      \methodt: Establishing Protein as a Native \\ Modality in Large Language Models
}

\affiliation[1]{Shanghai Artificial Intelligence Laboratory}
\affiliation[2]{Generative Symbolic Intelligence Lab (GenSI), Tsinghua University}
\affiliation[3]{Institute for AI Industry Research (AIR), Tsinghua University}

\abstract{
    \vspace{-1mm}

\input{sections/00abstract}
    \vspace{-6mm}
}

\correspondence{\href{mailto:zhouhao@air.tsinghua.edu.cn}{zhouhao@air.tsinghua.edu.cn}}

\projectpage{\url{https://amix-bio.github.io/AMix-2/}}

\begin{document}

    \maketitle
    
    \vspace{-16mm}

    \begin{figure}[H]
        \centering
        \includegraphics[width=0.87\linewidth]{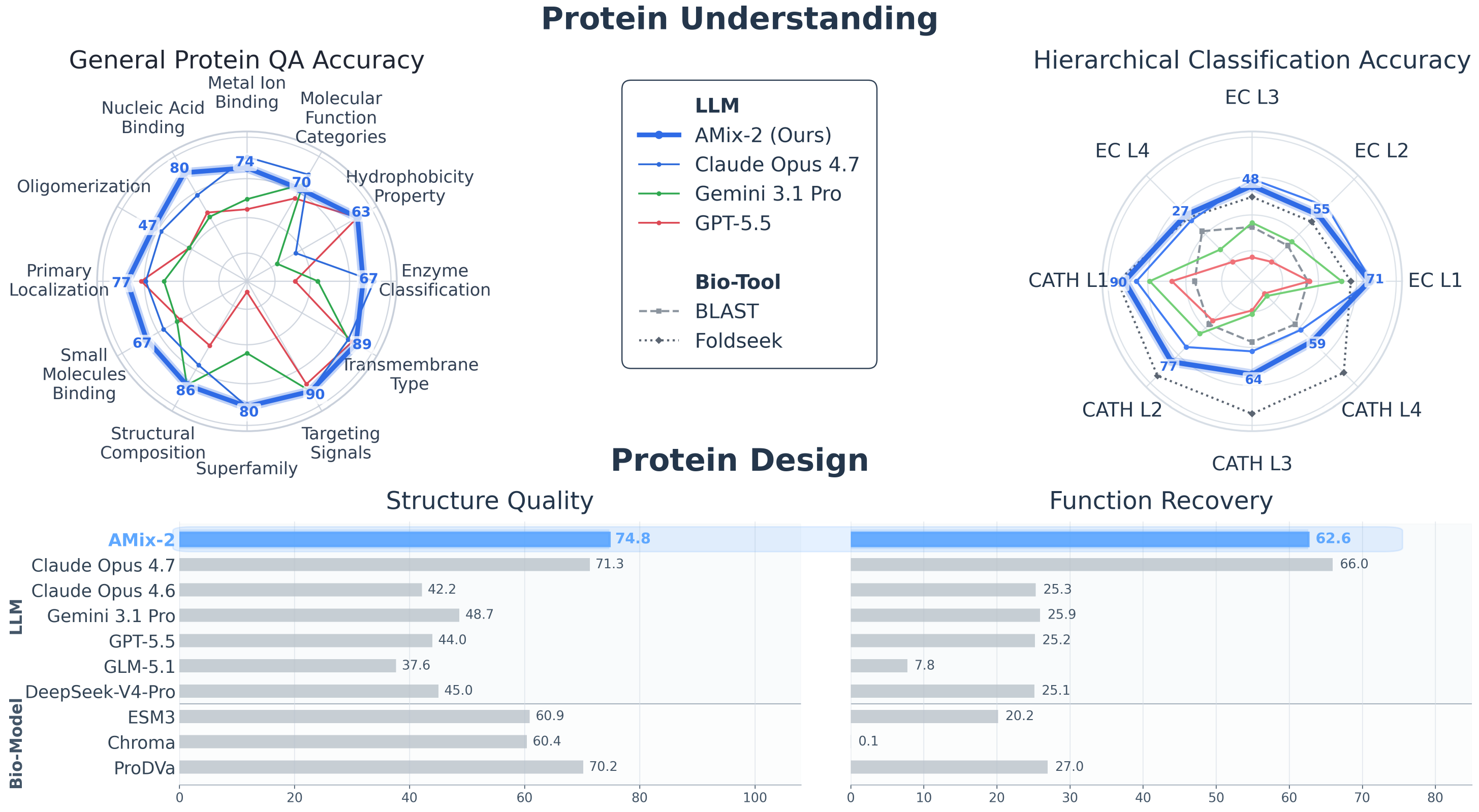}
        \caption{Performance overview of \method. \method outperforms frontier LLMs and demonstrates competitive performance to task-specific protein models on various protein understanding and design tasks.
        \vspace{-5mm}
        }
        \label{fig:overview}
    \end{figure}

    \newpage
    \tableofcontents
    \newpage







    \input{sections/01intro}

    \input{sections/02model}

\input{sections/03datasets}

    \input{sections/04experiments}

    \input{sections/05related}

    \input{sections/06conclusion}

    \input{sections/contributions}
    
    \clearpage
    \bibliographystyle{unsrtnat}
    \bibliography{refs}

    \clearpage
    \beginappendix
    \input{sections/appendix}

\end{document}

%% file: macro.tex
\usepackage{natbib}
\usepackage{latexsym}

\usepackage{url}
\usepackage{amssymb}
\usepackage[utf8]{inputenc}
\usepackage{microtype}
\usepackage{booktabs}
\usepackage{pifont} 
\usepackage{multirow}
\usepackage{makecell}
\usepackage{paralist}
\usepackage{xspace}
\usepackage{color}
\usepackage{xcolor}
\usepackage{colortbl}
\usepackage{adjustbox}
\usepackage{hyperref} 
\usepackage[edges]{forest}
\usepackage{tikz} 
\usepackage{caption}
\usepackage{amsfonts}
\usepackage{tcolorbox}
\usepackage{algorithm}
\usepackage{algpseudocode}
\usepackage{amsthm}

\usepackage{mathtools}
\usepackage[version=4]{mhchem}
\usepackage{array}  
\usepackage{graphicx}

\hypersetup{
    colorlinks,
    linkcolor={blue!80!black},
    citecolor={blue!80!black},
}
\tikzset{
    root/.style =             {align=center, text width=1cm, rounded corners=3pt, line width=0.3mm, fill=gray!10, draw=gray!80, font=\small},
    demographic/.style =         {align=center, text width=1.8cm, rounded corners=3pt, line width=0.3mm, fill=blue!10, draw=blue!80, font=\footnotesize},
    demographic_work/.style =    {align=center, text width=10cm, rounded corners=3pt, line width=0.3mm, fill=blue!10, draw=blue!0, font=\footnotesize},
    character/.style =         {align=center, text width=1.8cm, rounded corners=3pt, line width=0.3mm, fill=red!10, draw=red!80, font=\footnotesize},
    character_work/.style =    {align=center, text width=10cm, rounded corners=3pt, line width=0.3mm, fill=red!10, draw=red!0, font=\footnotesize},
    personalization/.style =           {align=center, text width=1.8cm, rounded corners=3pt, line width=0.3mm, fill=cyan!10, draw=cyan!80, font=\footnotesize},
    personalization_work/.style =      {align=center, text width=10cm, rounded corners=3pt, line width=0.3mm, fill=cyan!10, draw=cyan!0, font=\footnotesize},
    risk/.style =         {align=center, text width=1.8cm, rounded corners=3pt, line width=0.3mm, fill=orange!10, draw=orange!80, font=\footnotesize},
    risk_work/.style =    {align=center, text width=10cm, rounded corners=3pt, line width=0.3mm, fill=orange!10, draw=orange!0, font=\footnotesize},
}

\usepackage{CJK}


%% file: math_commands.tex
\usepackage{amsmath,amsfonts,bm}

\def\eqref#1{equation~\ref{#1}}

\def\1{\bm{1}}

\DeclareMathAlphabet{\mathsfit}{\encodingdefault}{\sfdefault}{m}{sl}
\SetMathAlphabet{\mathsfit}{bold}{\encodingdefault}{\sfdefault}{bx}{n}



%% file: sections/00abstract.tex
We present \method, a protein–text foundation model that establishes protein as a native modality in large language models (LLMs), unifying protein understanding and sequence design within a single foundation model. \method is built upon two key ideas: (1) a unified protein–text formulation that embeds natural language and protein sequence in a shared token space, enabling one model to perform biological reasoning and conditional design instead of separate downstream task-specialized models; and 
(2) a block-wise diffusion language modeling backbone that combines causal generation across blocks with bidirectional context and iterative refinement within blocks. 
This scheme better matches the intrinsic nature of proteins than a strict left-to-right factorization.
To evaluate protein foundation models under realistic generalization settings, we further introduce \bench, a comprehensive benchmark with time-aware and homology-aware protocols across various understanding and design tasks, and with baselines covering classical bioinformatics tools, protein-specialized models and LLMs.
On \bench, \method outperforms frontier LLMs and demonstrates competitive performance to task-specific protein models. 
Controlled experiments further show that the diffusion-based paradigm generally surpasses its autoregressive counterpart, highlighting the advantage of flexible generation order for protein sequences. We release both \method and \bench to facilitate open research in protein foundation models.

%% file: sections/01intro.tex
\section{Introduction}

One of the next major frontiers for large language models (LLMs) is scientific discovery, with life science emerging as a particularly important domain.
Among biological entities, proteins are especially central: they carry out most cellular functions and are key objects for both scientific study and bioengineering \citep{jumper2021highly,doi:10.1126/science.abj8754,dauparas2022robust,watson2023novo,abramson2024accurate}. 
A model that can directly read, reason about, and even generate proteins would therefore mark a meaningful step toward more capable scientific foundation models.

Current progress towards this goal has followed two complementary 
paths: 
(1) Protein-specialized foundation models \citep{lin2023evolutionary,madani2023large,su2023saprot,su2024protrek,hayes2025simulating} and bioinformatics tools \citep{altschul1990basic,steinegger2017mmseqs2,van2024fast} have driven major advances in representation learning, structure prediction, sequence generation, and similarity-based search, but remain largely \emph{specialized and isolated}: limited to the protein modality, with interfaces specifically designed for certain fixed-format objectives such as embedding extraction or masked prediction, rather than open-ended language instruction. In practice, protein understanding and sequence design are often handled by different models or by separate task-specific fine-tuning pipelines \citep{ZHANG2025103066}. 
(2) Frontier LLM systems~\cite{anthropic2026claude4, google2026gemini3, singh2025openai, deepseek2026v4, zeng2026glm, yang2025qwen3}, on the other hand, are \emph{general and externally scaffolded}: they offer flexible natural-language interfaces and can coordinate biological workflows with harnessing. Equipped with retrieval, database access, and other tool use, such systems have shown strong promise for biological problem solving \citep{huang2025biomni,jin2025stella,fallahpour2025bioreason}. However, in many cases their protein capabilities are mediated primarily through external resources, rather than on protein knowledge learned natively within the model itself. In this work, we use \emph{native protein capability} to refer to the ability to understand and generate proteins directly from protein sequence inputs and instructions, without relying on tools or complex pipelines. 

We present \method, a protein--text foundation model that treats proteins as a native modality within a unified language-modeling framework, combining the protein-modeling capacity of specialist models with the instruction-following flexibility of large language models. 
In \Cref{sec:unified_modeling_data}, \method formulates protein understanding and functional sequence design as protein--text conditional generation in a shared token space: with natural-language instructions specifying the task, protein sequences serving as inputs or outputs, \method reasons about its functional context. This unified formulation allows a single model to perform diverse protein sequence understanding and design tasks through various instructions.
\Cref{sec:dllm_architecture} describes how \method uses a block-wise diffusion language modeling backbone. This choice is motivated by a mismatch between protein tasks and standard left-to-right generation. Protein sequences exhibit non-local dependencies \citep{rives2021biological,rao2021msa,rao2021transformer}, and many practical tasks involve partial infilling, local editing and iterative refinement rather than purely autoregressive continuation \citep{wang2024diffusion}. Block-wise diffusion addresses this by combining causal generation across blocks with bidirectional denoising within each block, supporting both global consistency and fine-grained controllability \citep{arriola2025block,song2025seed}. Under matched training data, our diffusion backbone outperforms its autoregressive counterpart with large improvements on protein design (\Cref{tab:design_performance_comparison}). 

To establish a rigorous, shared testbed for assessing native protein capabilities, we introduce \bench in \Cref{sec:proteinarena}, a benchmark that follows standard \emph{time-aware} and \emph{homology-aware} evaluation protocols across protein question answering, hierarchical enzyme commission number (EC) and structural fold (CATH) classification, and function-conditioned de novo design \citep{10.1093/nar/gkac1052,yang2024care,waman2025cath,kuang2025pdfbench}. These protocols are important for assessing realistic generalization across model families. Despite the fact that for frontier LLMs with broad and largely opaque pretraining corpora, temporal and homology exposure cannot be completely ruled out as the other models, \bench still provides an informative evaluation ground to assess them. By comparing frontier LLMs, protein language models, and specialist bioinformatics tools under the same setting, \bench benchmarks these methods in the broader landscape of protein modeling.

On \bench, \method generally outperforms non-controlled general-purpose LLMs and shows strong competitiveness with protein-specialized language models and dedicated bioinformatics tools under strict splits (\Cref{sec:experiment_results}). 
It achieves state-of-the-art accuracy in General Protein QA (65.70\%), generalizes well to low-homology data regimes in EC and CATH classification,
while simultaneously yielding high structural plausibility and function recovery rate in protein design.
Together, these results support unified protein--text modeling as an effective route to stronger native protein capability. \method encourages protein knowledge to be internalized within the model and transferred across tasks, rather than fragmenting it into separate task-specific models and engineering pipelines.

We summarize our contributions as follows:
\begin{itemize}
\item We introduce \method, a protein--text foundation model with a block-wise diffusion language modeling backbone that supports both global constraints and local refinement.
\item We introduce \bench, a benchmark with strict evaluation protocols across a wide range of protein understanding and design tasks, providing a shared testbed for different model families.
\item We show that \method outperforms frontier LLMs, protein language models, and specialist bioinformatics tools on \bench, and that our dLLMs backbone demonstrates its superiority in protein modeling.
\end{itemize}

%% file: sections/02model.tex

\section{Unified Protein-Text Modeling}
\label{sec:unified_modeling_data}

Our goal is to treat proteins as a native modality within a single foundation model. Rather than building separate architectures or task-specific interfaces for protein understanding and sequence design, we cast both as instruction-following tasks over mixed text and protein sequences. This section presents the unified modeling interface and the pipeline used to construct large-scale training data under this formulation.

\subsection{Task Formulation}
\label{subsec:unified_token_space}

\method establishes a single discrete vocabulary $\mathcal{V}$ spanning both natural language and protein amino acids, enabling heterogeneous modalities to be processed within a shared representation space. Concretely, text is tokenized with a standard subword tokenizer and protein sequences are tokenized at the residue level. 
Each example is serialized as a single sequence that interleaves text tokens $x_{\text{text}}$ and protein tokens $x_{\text{prot}}$, so that both modalities are processed by one model under a unified instruction-following objective.

The two main task families differ only in the direction of generation. For \emph{protein understanding}, the model reads a text instruction and a protein sequence, then generates a text answer:
\begin{equation}
p\!\left(x_{\text{text}}^{\text{ans}} \;\middle|\; x_{\text{text}}^{\text{inst}},\, x_{\text{prot}}\right).
\end{equation}
As for \emph{sequence design}, the model reads a text instruction and generates a protein sequence:
\begin{equation}
p\!\left(x_{\text{prot}}^{\text{ans}} \;\middle|\; x_{\text{text}}^{\text{inst}}\right).
\end{equation}

Protein understanding refers to tasks in which the model reads a sequence and answers a textual query, such as a functional description, class label, localization or other property prediction regarding specific aspects of the protein. 
Sequence design refers to tasks in which the model takes a textual specification of desired functional or biochemical properties, and is required to translate the specified functional requirements to a novel protein sequence.
By expressing both tasks as instruction-following in a shared token space, \method unifies protein understanding and sequence generation within a single modeling interface.
Unlike previous protein generative models trained solely on protein data with fixed conditional inputs such as caption labels, which typically assumes a fixed conditioning direction (e.g.\ text $\rightarrow$ protein) and relies on modality-specific fusion mechanisms \citep{pei2025leveraging}, or embedding-only models such as ESM2~\cite{lin2023evolutionary} designed for representation tasks, this formulation is symmetric: the model can condition on and generate either text or protein tokens, enabling a more general \emph{any-to-any} modeling interface across modalities and tasks.

\subsection{Data Construction}
\label{subsec:data_construction}

To support the unified protein--text formulation, we construct a multimodal corpus that expresses diverse biological knowledge from curated biological databases including UniProtKB~\cite{10.1093/nar/gkac1052}, UniRef50~\cite{steinegger2018clustering}, InterPro~\cite{blum2025interpro} and CARE~\cite{yang2024care}.
Training proceeds in two stages: \textbf{continual pre-training} for protein knowledge injection and \textbf{post-training} for downstream task alignment.

\paragraph{Continual Pre-Training}
We pair the protein sequences of UniRef50 with textual descriptions derived from UniProtKB entries, together with general-domain text to preserve broad language competence. The goal of this stage is to 
introduce protein sequences into the model's pretraining distribution,
so that amino-acid patterns and sequence-level regularities are learned jointly with natural-language semantics rather than treated as out-of-distribution character strings.
This stage equips \method with broad aligned knowledge over proteins and text, providing the foundation for downstream protein understanding and design.

\paragraph{Post-Training}
We build an instruction-following dataset spanning both protein understanding and sequence design. Protein question answering examples are constructed from Swiss-Prot, a manually reviewed subset of UniProtKB \citep{10.1093/nar/gkg095}. Function-conditioned \textit{de novo} design cases are derived from InterPro functional keywords and motif descriptions. Hierarchical enzyme commission (EC) and CATH structural fold classification tasks are built from CARE and curated Swiss-Prot annotations, respectively. Each example is formatted as an instruction, the relevant sequence or functional context, and a target output. We additionally attach rationales derived from protein records where appropriate to further boost the reasoning capabilities of \method. These rationales undergo iterative filtering for \textit{factual accuracy}, \textit{logical integrity} and \textit{context consistency}.

Together, the two-stage pipeline transforms heterogeneous biological resources into a unified protein--text training corpus, enabling an all-in-one model of \textbf{comprehension}, \textbf{classification}, and \textbf{sequence design} through instructions alone. Notably, sequences with high similarity to test proteins, or with release dates after the temporal cutoff, have been removed from the labeled training sets for data decontamination.

    \begin{figure}[t]
        \centering
        \includegraphics[width=\linewidth]{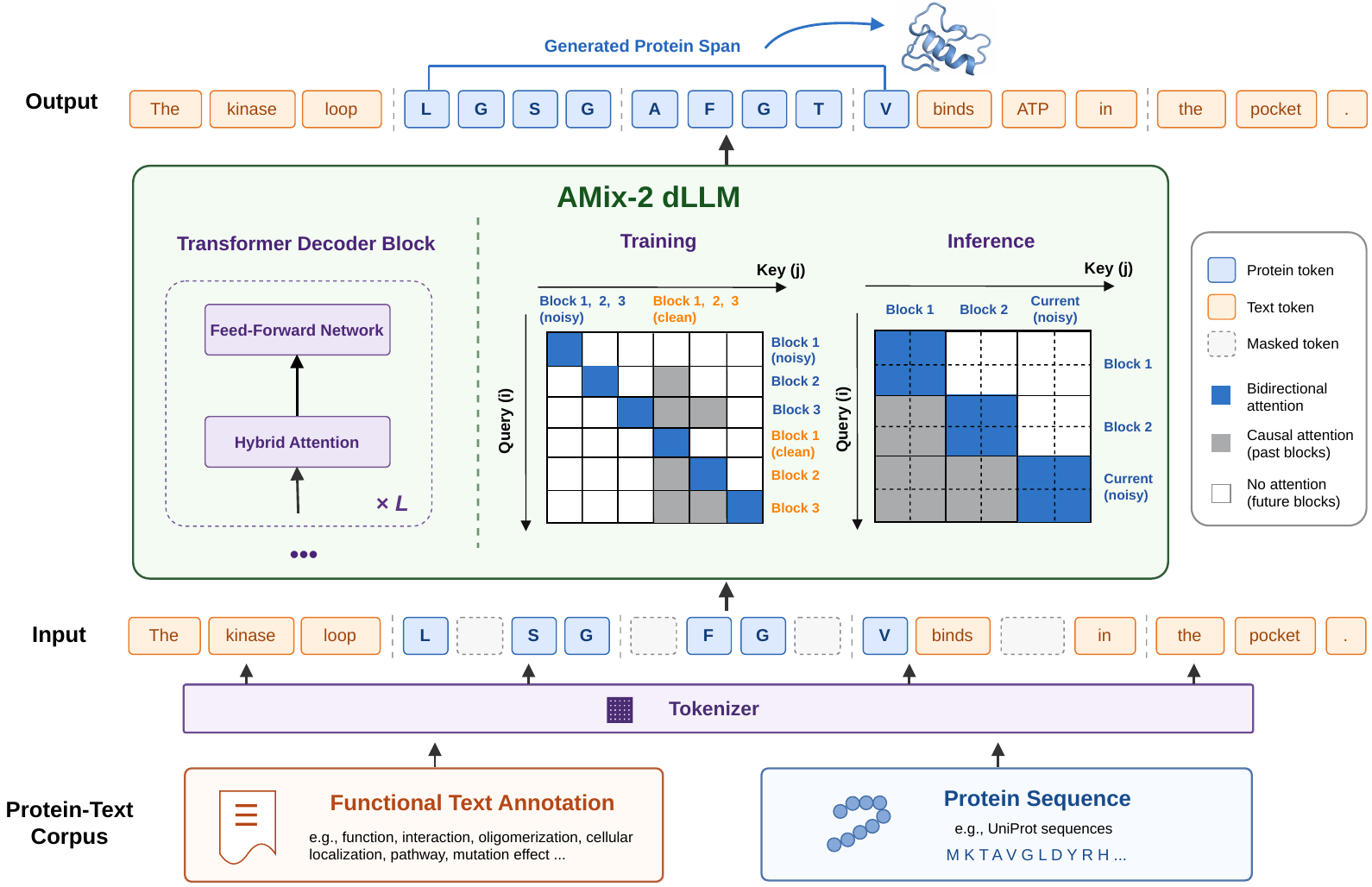}
        \caption{Block-wise diffusion LLM architecture of \method, jointly reasoning over the unified protein--text token space. Protein sequences and functional text annotations are tokenized into a single mixed sequence, where a subset of tokens are masked and predicted according to a noise schedule. The model is implemented as a decoder-only Transformer with hybrid attention: during training, tokens within each diffusion block attend bidirectionally to one another while also attending causally to previous blocks; during inference, previously denoised blocks are treated as fixed context and stored in KV Cache, and the current noisy block is iteratively refined.}
        \label{fig:dllm_arch}
    \end{figure}

\section{Block-wise Diffusion LLM Architecture}
\label{sec:dllm_architecture}


Given the unified protein--text interface, we instantiate \method as a \emph{block-wise diffusion large language model} (dLLM).
The key motivation is that protein--text joint modeling is not always well served by token-level autoregressive (AR) decoding: while text may be sequential, protein residues naturally exhibit long-range, non-local dependencies, making the standard AR next-token prediction objective less suitable for practical tasks such as motif scaffolding, local editing, and iterative refinement under global constraints.
To better accommodate this mixed-modality dependency,
\method preserves \emph{causal ordering across semantic blocks} for global coherence, while using \emph{bidirectional masked diffusion within each block} to refine tokens jointly. This yields a unified architecture that retains LLM-style controllability over mixed protein--text sequences while better supporting dependency patterns and editing requirements common in protein tasks.


\subsection{Mask-based Discrete Diffusion}
\label{subsec:mask_diffusion}

\method models the generative process using mask-absorbing discrete diffusion~\citep{austin2021structured,wang2024diffusion,nie2025large} over the shared discrete vocabulary $\mathcal{V}$ of size $K$, covering both text sub-words and protein residues.
Augmenting $\mathcal{V}$ with a special \texttt{[MASK]} state,  each clean token is represented by a one-hot vector $\mathbf{x}_0 \in \{e_1,\ldots,e_K\} \subset \mathbb{R}^{K+1}$, and let $\mathbf{m} = e_{K+1}$ denote the absorbing \texttt{[MASK]} token. 
\paragraph{Forward process}
The forward state at step $t \in [0,1]$ is governed by the underlying transition matrix
\begin{equation}
    Q_t \;=\; \alpha_t\,I \;+\; (1-\alpha_t)\,\mathbf{1}\mathbf{m}^{\top}
    \;\in\; \mathbb{R}^{(K+1)\times(K+1)},
    \label{eq:transition_matrix}
\end{equation}
where $\alpha_t \in [0,1]$ is a monotonically decreasing noise schedule.
Under this kernel, each token remains unchanged with probability $\alpha_t$ and is absorbed into \texttt{[MASK]} with probability $1 - \alpha_t$; once masked, a token stays masked for all subsequent times.
The resulting marginal forward distribution is
\begin{equation}
    q(\tilde{x}_t \mid x_0) = \text{Categorical}(\tilde{x}_t; \alpha_t x_0 + (1-\alpha_t)\mathbf{m})
    \label{eq:marginal}
\end{equation}
\paragraph{Reverse process}
A neural network $p_\theta$ is trained to predict the clean sequence $\mathbf{x}_0$ from the corrupted sequence $\tilde{\mathbf{x}}_t$, parameterizing the reverse denoising
process by approximating the reverse posterior transition
\begin{equation}
    q(\tilde{x}_s \mid \tilde{x}_t, x_0) = 
    \begin{cases} 
    \text{Categorical}(\tilde{x}_s; x_0) & \text{if } \tilde{x}_t \neq \mathbf{m} \\[8pt] 
    \text{Categorical}\left(\tilde{x}_s; \dfrac{\alpha_s - \alpha_t}{1-\alpha_t}x_0 + \dfrac{1-\alpha_s}{1-\alpha_t}\mathbf{m}\right) & \text{if } \tilde{x}_t = \mathbf{m} 
    \end{cases}
    \label{eq:reverse_posterior}
\end{equation}
This mask-absorbing formulation is particularly well suited to protein modeling because it naturally supports partial observation, global conditioning, and iterative refinement.


\subsection{Block-wise Factorization}
\label{subsec:block_factorization}
A pure masked-diffusion model denoises the all tokens in parallel. Although this provides maximal flexibility in generation order, it makes variable-length generation less straightforward \citep{sahoo2024simple,li2025beyond} and leaves the model to resolve unnecessary uncertainty by treating all positions as simultaneously unconstrained. 
In contrast, \method imposes a \emph{block-level} autoregressive factorization on top of the diffusion process \citep{arriola2025block}. Blocks are generated causally, while tokens within each block are denoised bidirectionally, preserving informative prefix conditioning without sacrificing the flexibility of intra-block arbitrary-order generation.

The interleaved sequence $\mathbf{x}$ is partitioned into $L$ contiguous blocks of block size $D$,
\begin{equation}
\mathbf{x} = (B_1, B_2, \ldots, B_L).\label{eq:block_partition}
\end{equation}
\method factorizes the joint distribution at the block level:
\begin{equation}
p(\mathbf{x}) = \prod_{k=1}^{L} p(B_k \mid B_{<k}),\label{eq:block_factorization}
\end{equation}
where $B_{<k} = (B_1,\ldots,B_{k-1})$ denotes the preceding clean or fully denoised blocks.
The corrupted block at noise level $t$ is obtained by independently masking each token:
\begin{equation}
    q\!\left(\widetilde{B}_{k,t} \mid B_k\right)
    \;=\;
    \prod_{i \in B_k}
    q\!\left(\tilde{x}_{i,t} \mid x_{i,0}\right).
    \label{eq:block_corruption}
\end{equation}

The key difference from a standard autoregressive model is that each conditional distribution $p(B_k \mid B_{<k})$ is not generated from left to right. Instead, all $D$ tokens of $B_k$ are recovered {jointly} via masked diffusion
conditioned on $B_{<k}$, enabling bidirectional interactions within the block. 
This hybrid factorization preserves contextual conditioning across blocks while avoiding a rigid within-block left-to-right ordering that is often poorly matched to protein structure and function \citep{Alamdari2023.09.11.556673}.

\subsection{Parallel Training Objective and Hybrid Attention}
\label{subsec:training_objective}
\paragraph{Training Objective}
During training, we sample a block index $k$ and a diffusion time $t$, corrupt the current block to obtain $\widetilde{B}_{k,t}$, and train the model to recover the clean tokens in the masked positions. The resulting objective is
\begin{equation}
\mathcal{L}_{\text{block}}(\theta)
=
\mathbb{E}_{\mathbf{x},k,t,\widetilde{B}_{k,t}}
\left[
w(t)
\sum_{i \in M_{k,t}}
-\log p_{\theta}\!\left(x_{i,0} \mid B_{<k}, \widetilde{B}_{k,t}\right)
\right],
\label{eq:block_objective}
\end{equation}
where $M_{k,t}$ denotes the masked positions in the target block, and $w(t)$ weights different noise levels.
The ELBO can be reduced to a weighted cross-entropy over the masked positions \citep{sahoo2024simple, shi2024simplified,ou2024absorbingdiscretediffusionsecretly}.

Compared with the standard AR objective
$\mathcal{L}_{\mathrm{AR}} = -\mathbb{E}\sum_i \log p_\theta(x_i \mid \mathbf{x}_{<i})$,
the block diffusion objective provides \emph{denser supervision}: at each training step,
the model is trained to reconstruct multiple masked positions simultaneously, so gradient signals are aggregated from all $|M_{k,t}|$ targets rather than a single next-token prediction.
This is particularly beneficial for protein modeling,
where residue identities are often correlated, and the model is required to capture the co-evolutionary constraints across multiple sites \citep{rives2021biological,10.1371/journal.pone.0028766}.

\paragraph{Hybrid Causal-Bidirectional Attention}
The block factorization in \Cref{eq:block_factorization} induces a structured attention mask that is the architectural centerpiece of \method, as illustrated in \Cref{fig:dllm_arch}.
At inference time, blocks are generated sequentially, and the current block is denoised conditioned on previously generated clean blocks. 
For efficient training, we implement the same dependency structure with a parallel masking scheme. Specifically, we pack all corrupted blocks together with their clean counterparts into a single sequence. 
\begin{equation}
    S \;=\;
    \bigl(
        \widetilde{B}_{1,t},\; 
        \widetilde{B}_{2,t},\; 
        \ldots,\;
        \widetilde{B}_{k,t},\; 
        \ldots,
        B_1,\; B_2,\;, \ldots, B_k,\;
        \ldots
    \bigr).
    \label{eq:training_sequence}
\end{equation}
For a query position $u \in \widetilde{B}_{k,t}$, the attention mask is defined so that $u$ may attend to all tokens in the same corrupted block and to all clean prefix blocks:
\begin{equation}
    A_{uv} = 1
    \iff
    \bigl(\underbrace{v \in \widetilde{B}_{k,t}}_{\text{bidirectional}}\bigr)
    \;\;\text{or}\;\;
    \bigl(\ \underbrace{v \in B_{<k}}_{\text{causal}}\ \bigr).
    \label{eq:attention_mask}
\end{equation}
Thus, each noisy block attends only to itself and the appropriate clean prefix that would be available at inference time, while all blockwise denoising losses can still be computed and optimized in parallel during a single forward pass. This makes the model well suited to protein--text tasks that require autoregressive conditioning on informative upstream content while refining protein tokens within a coherent local region.

%% file: sections/03datasets.tex
\section{ProteinArena: A Time-aware and Homology-aware Benchmark}
\label{sec:proteinarena}

Reliable evaluation of native protein capability requires more than random train--test splits. Protein datasets are highly redundant, and models can achieve deceptively strong performance by exploiting close homologs rather than learning transferable biochemical regularities \citep{NEURIPS2019_TAPE,dallago2021flip,NEURIPS2022_PEER}. In addition, for models trained on broad scientific or web-scale corpora, temporal leakage can confound evaluation if test proteins were already publicly available during training. To address these issues, we introduce \bench, a benchmark designed for fair evaluation of native protein understanding and functional sequence design.

\subsection{Benchmark Principles and Split Protocols}
\label{subsec:benchmark_protocol}

\bench is built around two principles: \emph{time awareness} and \emph{homology awareness}, partitioning training and evaluation data based on both temporal cutoffs and sequence identity thresholds. 

By retaining only proteins with a Swiss-Prot first public date on or after \textbf{January 1, 2025}, \bench removes the risk of temporal contamination for \method, as well as bioinformatics tools and protein language model baselines, which were all trained on sequences prior to this cut-off date. 
Moreover, our benchmark explicitly controls sequence similarity between training and evaluation sets. Under the primary test setting, only those proteins that exhibit less than 30\% sequence identity to any sequence released on or before \textbf{December 31, 2024} are included.
This low-homology regime is intended to measure genuine generalization rather than near-neighbor retrieval. We additionally report higher-homology ranges as supplementary analysis in Appendix~\ref{app:subsec:homology_identity}, but focus our main interest on the more challenging low-homology setting.



As \bench is designed to evaluate the native capabilities of models, assessment is conducted without external retrieval, database lookup, or workflow orchestration, thus keeping the focus of evaluation on model-internal protein competence.

\subsection{Protein Understanding}
\label{subsec:native_understanding}

The understanding track evaluates newly released, low-homology proteins using a variety of tasks spanning both open-form question answering and fine-grained classification, measuring fundamental aspects of protein understanding such as function, interaction, localization and structure.

\paragraph{General Protein QA}
To evaluate open-ended protein understanding in a natural-language setting, we construct a General Protein QA subset from reviewed Swiss-Prot entries released under the benchmark protocol. 
The objective of this task is to measure whether a model can infer biologically meaningful properties directly from sequence and express them through text, rather than only producing fixed labels.
As illustrated in \Cref{fig:general_qa_benchmark}, this subset contains 481 samples across 16 representative categories, including cleavage sites, enzyme classification, functional domains, hydrophobicity property, molecular function categories, metal ion binding, nucleic acid binding, oligomerization, post-translational modifications, primary localization, protein family, small molecules binding, structural composition, superfamily, targeting signals, and transmembrane type. Each sample pairs a protein entry with a paraphrased natural-language question, providing the ability to probe whether a model can infer biologically meaningful properties directly from the sequences given. Instruction instances for each category are detailed in Appendix~\ref{app:subsec:general_protein_qa_data}.

\paragraph{Hierarchical Classification}
To facilitate direct comparison with protein-specialized models and classical bioinformatics tools, \bench further includes hierarchical classification tasks of EC and CATH numbers. These tasks require the evaluated model to predict and assign labels across all four levels of the corresponding ontology, from Level 1 (\texttt{x.-.-.-}) to Level 4 (\texttt{x.x.x.x}), thereby testing progressively fine-grained functional and structural discrimination capabilities. Compared with coarse binary or family-level prediction, the four levels of EC and CATH classification provide a more stringent measure of biological understanding, evaluating whether a model can move beyond superficial sequence similarity and recover functionally and structurally meaningful information under strict generalization constraints.

\subsection{Protein Design}
\label{subsec:native_design}

The design track evaluates function-conditioned de novo sequence generation. Following recent functional protein design benchmark \cite{kuang2025pdfbench}, we construct prompts from expert-reviewed InterPro functional keywords recorded in Swiss-Prot annotations under the time-aware protocol. Given these functional constraints, the model conducts \textit{de novo} design, i.e., generating a protein sequence from scratch.

This task aims to evaluate the model's instruction-following capability in translating functional constraints into valid protein sequences. Therefore, our evaluation framework fundamentally assesses two complementary criteria: (1) \textit{functional alignment}, i.e., whether the generated sequence realizes the requested function, and (2) \textit{biophysical plausibility}, i.e., whether it resembles viable natural proteins in structural quality and sequence-level characteristics. We additionally track uniqueness and novelty to distinguish meaningful design from degenerate generation or simple memorization.
Evaluation metrics are further detailed in \Cref{sec:experiment_baselines}.



By jointly covering comprehensive understanding and function-conditioned design, \bench provides a unified benchmark for native protein capabilities, enabling direct comparison across frontier LLMs, protein language models, and specialist bioinformatics tools under a common, leakage-aware evaluation protocol.

%% file: sections/04experiments.tex
\section{Experiments}
\label{sec:experiment_results}

\subsection{Training Setup}
\label{sec:experiment_settings}

We train two architectural variants of \method—the block-wise diffusion model (dLLM) and the autoregressive model (AR)—following a standardized two-stage large language model training paradigm. Both models are initialized from the \texttt{Qwen3-4B-Base}~\cite{yang2025qwen3} backbone and undergo initial continual pre-training on large-scale protein--text corpora, followed by post-training on a curated mixture of protein understanding and design instruction data. A fixed evaluation split of 1,000 packed samples is reserved to monitor convergence. Across both architectures, training is conducted in mixed-precision (BF16) using the AdamW optimizer with weight decay of $0.01$, $\beta_2=0.995$, and gradient norm clipping at $1.0$. To enhance stability and generalization, an Exponential Moving Average (EMA) is applied to model weights with a decay rate of $0.9995$. Learning rates follow a cosine decay schedule with a peak of $5\times10^{-5}$.

\paragraph{\method~dLLM}
Serving as our flagship architecture, this model utilizes a block diffusion language modeling objective with a linear noise schedule and a block size $D=32$. For large-scale distributed training, we employ a global batch size of 768, pack samples to a length of 4096, and set the warmup ratio to $0.05$. Notably, for the continual pre-training phase, we employ a block size warming-up schedule for smoother gradient norms and more stabilized training, while post-training utilizes document-aware packing with sample boundaries aligned to the block size $D$. This diffusion-based approach allows for bidirectional refinement within each block, facilitating the global and local conditioning required for complex protein tasks.


\paragraph{\method~AR}
As a comparative baseline, the autoregressive variant is trained using a standard next-token prediction objective. To ensure a rigorous comparison, we maintain the same core hyperparameter configurations and training data as \method~dLLM, with the exception of using a global batch size of 256, a packing length of 8192, and a warmup ratio of $0.1$.

\subsection{Baselines and Metrics}
\label{sec:experiment_baselines}

We compare the dLLM and autoregressive backbone variants of our \method model against a suite of baselines: (1) \emph{Frontier LLMs} including Claude Opus 4.7~\cite{anthropic2026claude4}, Claude Opus 4.6~\cite{anthropic2026claude4}, Gemini 3.1 Pro~\cite{google2026gemini3}, GPT-5.5~\cite{singh2025openai}, DeepSeek-V4-Pro~\cite{deepseek2026v4}, GLM-5.1~\cite{zeng2026glm} and Qwen3.5-27B~\cite{yang2025qwen3}.
(2) For protein understanding, we compare against \emph{protein-specific representation models}, including ESM2~\cite{lin2023evolutionary}, ESM3~\cite{hayes2025simulating}, ProTrek~\cite{su2024protrek} and SaProt~\cite{su2023saprot}, alongside \emph{retrieval-based bioinformatics tools} such as BLAST~\cite{altschul1990basic} and Foldseek~\cite{van2024fast}.
(3) For sequence design, we consider baselines that are compatible with our conditioning setup, including both \emph{sequence-based} such as ProteinDT~\cite{liu2023text} and \emph{structure-based} such as Chroma~\cite{ingraham2023illuminating}, together with the multimodal ESM3, as detailed in Appendix~\ref{app:subsec:baseline}.
Notably, structure-based models such as RFdiffusion \citep{watson2023novo} are not included because they do not support the programmable functional-specification, and therefore are not directly comparable.

Crucially, while frontier LLMs typically do not restrict their training data regarding specific protein sequences, all included protein-specific models and bioinformatics tools were verified to have training and reference databases cutoff dates prior to \textbf{December 31, 2024}. This ensures that no overlap exists between their supervised training samples and the test in \bench, consistent with the setting used for \method.

Specifically, for the \textbf{General Protein QA} task, we employ LLM-as-a-judge~\cite{gu2025surveyllmasajudge} using Gemini 3 Flash, evaluating whether the model-generated answers are semantically identical to the ground-truth protein functions. 
Traditional protein language models and specialist bioinformatics tools are omitted from this task, as their architectures are designed for representation learning or classification rather than free-form answer generation in a conversational QA-based setting.
For the \textbf{Functional De Novo Design} task, 
following the evaluation setup of \citet{kuang2025pdfbench}, 
we assess the generated sequences using a comprehensive suite of structural and functional metrics. Structural plausibility is measured via the ESMFold~\cite{lin2023evolutionary} predicted Local Distance Difference Test (pLDDT). To evaluate functional adherence, we compute Function Recovery (InterPro Recovery, IPR) by comparing the functional labels—obtained by scanning the generated sequences with InterProScan—against the ground-truth annotations. Furthermore, we quantify the sequence-level and structure-level novelty and diversity of the generated proteins utilizing MMseqs2~\cite{steinegger2017mmseqs2} and Foldseek, respectively. 
For the hierarchical classification tasks of \textbf{EC Prediction} and \textbf{CATH Prediction}, sequences are first grouped by identity, and performance is then measured by calculating accuracy metrics according to the precise correctness of predicted labels across the four hierarchical levels, for example ``5.6.2.4''. Details of metrics are left to Appendix~\ref{app:subsec:eval_setting}.

\subsection{Results and Analysis}
\label{subsec:experiment_results}

\begin{figure}[tb]
    \centering
    \includegraphics[width=1\linewidth]{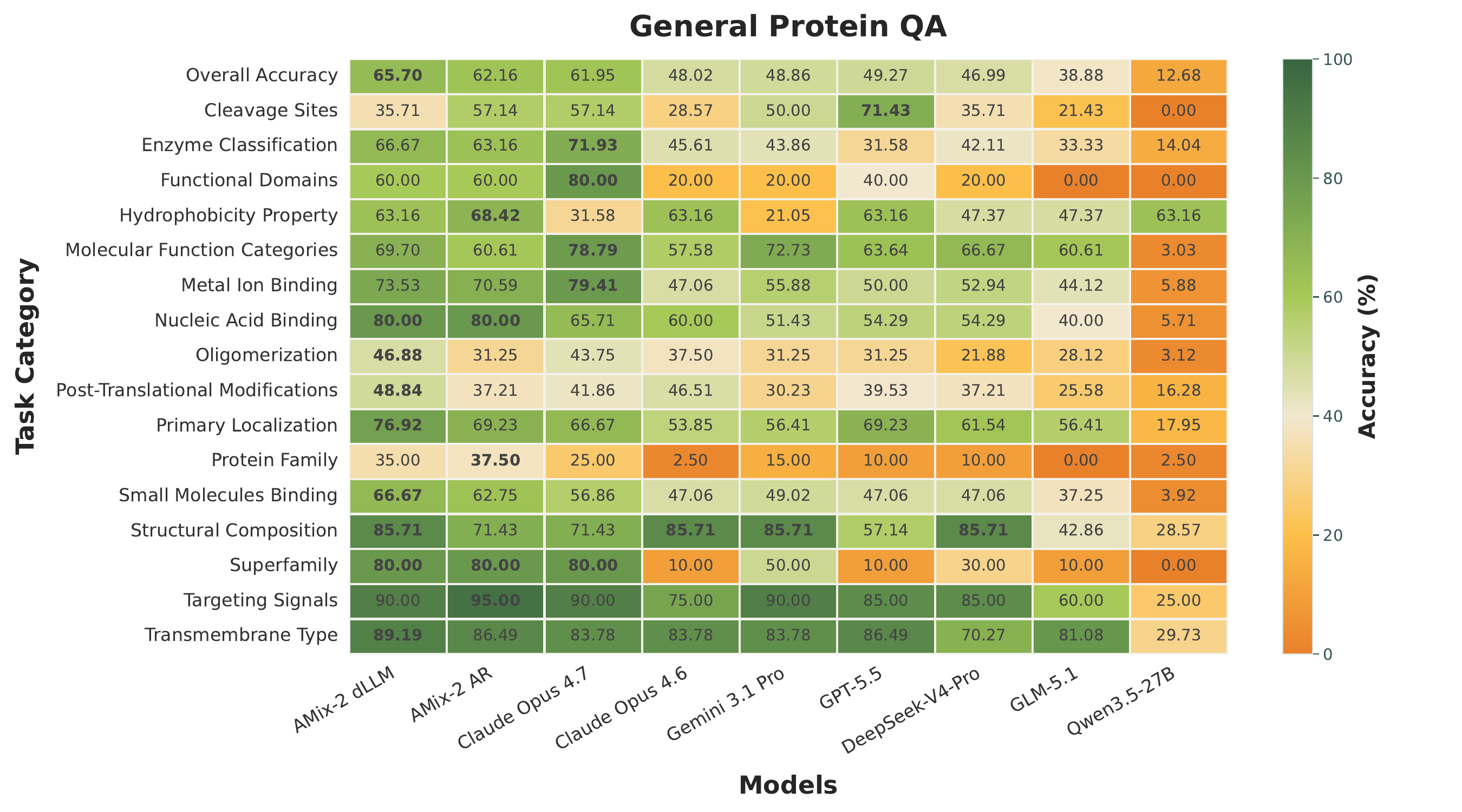}
    \caption{General Protein QA performance across \method and frontier large language models.}
    \label{fig:model_performance_heatmap_wrapped}
\end{figure}

\paragraph{General Protein QA} 
We evaluate the performance of our proposed model, \method, on \bench across 16 distinct protein functionality categories, comparing it against state-of-the-art frontier models. To ensure rigorous assessment, all proteins sharing a sequence identity greater than 30\% with \bench were strictly excluded from the training of \method. As illustrated in \Cref{fig:model_performance_heatmap_wrapped}, the \method~dLLM variant achieves the highest overall accuracy of 65.70\%, followed by \method~AR at 62.16\%. Both variants surpass the leading frontier model Claude Opus 4.7 at 61.95\%, and substantially outperform other gigantic general-purpose LLM baselines such as GPT-5.5 and Gemini 3.1 Pro. While particular models exhibit higher accuracy at tasks like Cleavage Sites and Functional Domains, both variants of \method remain highly competitive, outperforming the majority of the frontier cohort.
Detailed task-specific analysis yields the following insights:

\begin{itemize}
    \item \textbf{Taxonomic and Evolutionary Classification:} In Protein Family and Superfamily classification, \method achieves an accuracy of 35\% and 80\% respectively, whereas leading models like Claude Opus 4.6, Gemini 3.1 Pro and GPT-5.5 struggle significantly. This highlights \method's ability to map sequences to evolutionary clusters even in low-homology regimes.

    \item \textbf{Structural and Localization Properties:} Our \method models show exceptional ability in predicting structural and localization properties, achieving near-ceiling performance on the tasks of Targeting Signals, Transmembrane Type, and Structural Composition. 

    \item \textbf{Interaction and Binding:} \method leads the benchmark in tasks such as Small Molecules Binding, Oligomerization, and Nucleic Acid Binding, while remaining competitive on Metal Ion Binding, demonstrating its ability to recognize protein interaction and binding types. 
\end{itemize}

\paragraph{Functional De Novo Design}
Table~\ref{tab:design_performance_comparison} shows \method~dLLM effectively integrates the strengths of protein-native modeling with the semantic reasoning capabilities of LLM foundations. It achieves superior generation quality, yielding both high pLDDT and functional recovery rate (IPR), significantly exceeding the vast majority of evaluated frontier LLMs in protein quality and surpassing specialized bio-models in satisfying text-specified functional constraints. 
We summarize our findings from the design results below:

\begin{itemize}
    \item \textbf{Design Failure in General LLMs:} Most off-the-shelf frontier models exhibit substantial weaknesses in protein design, typically suffering from pathological n-gram sequence repetition and structural validity. For instance, Qwen3.5 yields a Rep5 score of 63.24, while most models produce poor structural quality with pLDDT scores below 50. A notable exception is Claude Opus 4.7, showing highly competitive performance for protein sequence modeling compared to its peers.

    \item \textbf{Usability Bottlenecks of Specialist Models:} Protein-specialized models generally produce more structurally plausible sequences, but their practical utility is limited by rigid conditioning mechanisms. Unlike natural-language interface for LLMs, these models require task-specific input adaptations tied to their training formats--ranging from raw InterPro IDs and mapped class labels to exact UniProt text templates, greatly restricting their flexibility (exemplified in \Cref{tab:bio_model_conditioning_examples}). Notably, CFPGen yields valid samples for only 36.7\% of the test cases. To enable a better controlled apples-to-apples comparison, we report results on the subset shared across all baselines in Appendix~\ref{app:subsec:homology_identity}.
    
    \item \textbf{Advantage of the dLLM Architecture:} Our results indicate that dLLM architecture is inherently better suited for protein design than traditional autoregressive counterparts. Compared with our standard \method~AR variant, the dLLM architecture provides a substantial boost in sequence quality and effectively mitigates the degeneration and overfitting behavior observed in AR models for such data-constrained scenarios, yielding higher diversity and substantially improved uniqueness.
\end{itemize}

\begin{table}[t]
\centering
\small
\renewcommand{\arraystretch}{1.05}

\begin{tabular}{l >{\rmfamily}l 
SSS SS SS
} 
\toprule
    \multirow{2}{*}{\textbf{Category}}
    & \multirow{2}{*}{\textbf{Model}}
    & \multicolumn{3}{c}{\textbf{Repetition}}
    & \multicolumn{2}{c}{\textbf{Quality}}
    & \multicolumn{2}{c}{\textbf{Seq Distribution}} \\
\cmidrule(lr){3-5} \cmidrule(lr){6-7} \cmidrule(lr){8-9}
    & 
    & \textbf{Rep} & \textbf{Rep2} & \textbf{Rep5} & \textbf{pLDDT} & \textbf{IPR (\%)} & \textbf{Novelty} & \textbf{Unique (\%)} \\
\midrule

\addlinespace[0.1em]
\multirow{1}{*}{\textbf{Reference}}
& Natural & 2.23 & 44.05 & 0.48 & 75.75 & 100.00 & 4.82 & 99.43 \\
\addlinespace[0.1em]
\midrule

\addlinespace[0.1em]
\multirow{6}{*}{\textbf{Bio-Model}}
& CFPGen & 13.52 & 60.15 & 15.77 & \ul{74.45} & 33.08 & 50.26 & 100.00 \\
& Chroma          &  2.68 & 51.92 &  0.47 & 60.38 &  0.05 & 59.56 & 100.00 \\
& ESM3 & 28.39 & 69.46 & 21.64 & 60.56 & 19.20 & 73.70 & 100.00 \\
& Pinal           & 14.28 & 52.11 &  4.68 & 68.49 & \ul{36.44} & 49.76 & 99.98 \\
& ProDVa          &  3.91 & 25.29 &  6.69 & 70.22 & 26.96 & 22.86 & 82.53 \\
& ProteinDT       &  5.09 & 52.25 &  2.16 & 39.53 &  8.98 & 61.05 & 99.90 \\
\addlinespace[0.1em]
\midrule
\addlinespace[0.5em]
\multirow{9}{*}{\textbf{LLM}}
& Qwen3.5-27B     & 15.98 & 76.95 & 63.24 & 33.47 &  0.47 & 77.93 & 100.00 \\
& GLM-5.1         &  1.55 & 47.34 &  4.32 & 37.64 &  7.75 & 54.46 & 100.00 \\
& DeepSeek‐V4‐Pro &  6.92 & 50.32 &  8.14 & 45.03 & 25.14 & 51.25 & 100.00 \\
& GPT‐5.5         &  2.99 & 53.19 &  3.86 & 43.95 & 25.21 & 44.21 & 100.00 \\
& Gemini 3.1 Pro  &  6.38 & 45.23 &  7.61 & 48.91 & 25.77 & 46.17 & 99.06 \\
& Claude Opus 4.6 &  2.24 & 48.44 &  1.86 & 42.17 & 25.32 & 45.91 & 97.59 \\
& Claude Opus 4.7 &  2.43 & 44.65 &  0.65 & 71.33 & \textbf{65.95} & 36.54 & 99.86 \\

\addlinespace[0.1em]
& \cc \textbf{\method~AR}   & \cc 6.99 & \cc 48.35 & \cc 6.3 & \cc 64.02 & \cc 39.07 & \cc 11.83 & \cc 49.68 \\
& \cc \textbf{\method~dLLM} & \cc 3.47 & \cc 45.22 & \cc 1.23 & \cc \textbf{74.75} & \cc 62.61 & \cc 30.57 & \cc 95.73 \\
\addlinespace[0.1em]

\bottomrule
\end{tabular}
\caption{Comparison of various models on generation quality, novelty, and sequence uniqueness metrics.}
\label{tab:design_performance_comparison}
\end{table}

\paragraph{Hierarchical Classification}
The performance of our \method model on the hierarchical classification of Enzyme Commission (EC) numbers and CATH structural domains is detailed in \Cref{tab:performance_ec_cath}, demonstrating \method's capability of capturing structural and functional features in low-homology regimes where traditional alignment tools like BLAST struggle. Notably, \method also surpasses frontier LLMs by a greater margin in high-homology regimes, as shown in Appendix~\ref{app:subsec:homology_identity}.

\begin{itemize}    
    \item \textbf{EC Prediction:} On the task of EC Prediction, \method~AR and \method~dLLM achieves leading Level-4 accuracies of 30.30\% and 27.27\%, while the other competitive specialized model ESM2 hits 28.28\% and Claude Opus 4.7 yields 25.25\%. \method significantly outperforms other general-purpose LLMs such as Gemini 3.1 Pro and DeepSeek-V4-Pro. This highlights how \method's hierarchical decomposition training strategy for EC numbers allows it to effectively bridge the gap between protein functions and fine-grained enzyme classifications through linguistic reasoning. 

    \item \textbf{CATH Prediction:} Within the general-purpose LLM category, \method~dLLM sets a new baseline for structural fold classification from sequence alone, achieving the best performance across all CATH levels, with \method~AR consistently ranking among the strongest LLM baselines as well. Both variants clearly outperform other frontier LLM baselines, highlighting the benefit of unified protein--language modeling for structural inference. While specialist bio-models and structure-aware tools still remain stronger overall, \method represents a meaningful step toward high-fidelity fold classification without requiring explicit 3D coordinate inputs, unlike methods such as Foldseek, SaProt, and ProTrek.

    \item \textbf{Domain-Specific Advantages vs. LLM Limitations Analysis:} 
    The stronger performance of protein-specialized methods is expected, as these tasks are closely tied to sequence patterns that encode protein function and structure.
    Classical bioinformatics tools directly leverage sequence similarity and homology-based transfer, and protein language models benefit from pretraining on large protein datasets that capture domain-specific biochemical representations. 
    General-purpose LLMs, in contrast, are primarily optimized for natural language and typically treat protein sequences as peripheral inputs rather than a native modeling target. By internalizing protein knowledge within a shared foundation model, \method narrows this gap: it retains the flexible language interface of LLMs while learning protein-aware representations that better support fine-grained functional and structural classification.
\end{itemize}

\begin{table}[t]
\centering
\small 
\renewcommand{\arraystretch}{1.05}

\begin{tabular}{l >{\rmfamily}l SSSS SSSS} 
\toprule
    \multirow{2}{*}{\textbf{Category}}
    & \multirow{2}{*}{\textbf{Model}}
    & \multicolumn{4}{c}{\textbf{EC Prediction}}
    & \multicolumn{4}{c}{\textbf{CATH Prediction}} \\
\cmidrule(lr){3-6} \cmidrule(lr){7-10} & 
    & \textbf{L1} & \textbf{L2} & \textbf{L3} & \textbf{L4} & \textbf{L1} & \textbf{L2} & \textbf{L3} & \textbf{L4} \\
\midrule

\addlinespace[0.3em]
\multirow{2}{*}{\textbf{Bio-Tool}}
& BLAST           & 36.36 & 31.31 & 29.29 & 21.21       & 44.07 & 44.07 & 44.07 & 44.07 \\
& Foldseek        & \ul{60.61} & \ul{49.49} & \ul{43.43} & \ul{26.26}       & \ul{94.07} & \ul{88.98} & \ul{88.14} & \ul{86.44} \\
\addlinespace[0.1em]
\midrule
\addlinespace[0.5em]
\multirow{4}{*}{\textbf{Bio-Model}}
& ESM2            & \ul{70.71} & \ul{57.58} & \ul{49.49} & \ul{28.28}       & 91.53 & \ul{88.14} & 85.59 & \ul{81.36} \\
& ESM3            & 59.60 & 44.44 & 39.39 & 18.18       & 93.22 & 85.59 & 78.81 & 73.73 \\
& ProTrek         & 62.63 & 45.45 & 44.44 & 24.24       & \ul{94.92} & \ul{88.14} & \ul{87.29} & \ul{81.36} \\
& SaProt          & 69.70 & 50.51 & 48.48 & 24.24       & 92.37 & 86.44 & 83.90 & 78.81 \\
\addlinespace[0.1em]
\midrule
\addlinespace[0.5em]
\multirow{9}{*}{\textbf{LLM}}
& Qwen3.5-27B     & 12.12 & 2.02  & 0.00  & 0.00        & 28.81 & 16.10 & 4.24  & 0.85  \\
& GLM-5.1         & 27.27 & 5.05  & 4.04  & 3.03        & 56.78 & 24.58 & 6.78  & 2.54  \\
& DeepSeek-V4-Pro & 45.45 & 19.19 & 18.18 & 7.07        & 66.95 & 44.07 & 27.97 & 20.34 \\
& GPT-5.5         & 37.37 & 18.18 & 14.14 & 9.09        & 59.32 & 40.68 & 22.88 & 14.41 \\
& Gemini 3.1 Pro  & 55.56 & 34.34 & 31.31 & 14.14       & 73.73 & 52.54 & 25.42 & 16.95 \\
& Claude Opus 4.6 & 42.42 & 21.21 & 18.18 & 9.09        & 64.41 & 38.14 & 16.95 & 14.41 \\
& Claude Opus 4.7 & 72.73 & \textbf{61.62} & 51.52 & 25.25       & 82.20 & 64.41 & 50.00 & 49.15 \\

\addlinespace[0.1em]
& \cc \textbf{\method~AR}   & \cc \textbf{76.77} & \cc 58.59 & \cc \textbf{54.55} & \cc \textbf{30.30}         & \cc {86.44} & \cc 69.49 & \cc {60.17} & \cc 56.78 \\

& \cc \textbf{\method~dLLM} & \cc 70.71 & \cc 54.55 & \cc 48.48 & \cc 27.27        & \cc \textbf{89.83} & \cc \textbf{77.12} & \cc \textbf{64.41} & \cc \textbf{59.32} \\
\addlinespace[0.1em]

\bottomrule
\end{tabular}
\caption{Level-wise EC and CATH prediction accuracy on proteins with $<30\%$ sequence identity.}
\label{tab:performance_ec_cath}
\end{table}

%% file: sections/05related.tex
\section{Related Work}

\paragraph{Protein-Specialized Foundation Models.}
Specialized protein language models (PLMs) have emerged as a powerful foundation for modeling protein sequences. For protein understanding, masked language models (MLMs) such as the ESM family \cite{rives2021biological,lin2022language,lin2023evolutionary}  learn informative representations through bidirectional context modeling, providing strong priors for downstream prediction tasks. However, as MLMs are optimized for reconstruction rather than direct generation, they are not naturally suited for \textit{de novo} sequence design.
For protein generation, autoregressive (AR) models like ProGen \cite{madani2020progen,nijkamp2023progen2,bhatnagar2026scaling} support sequence design but often underperform in representation learning. 
The diffusion-based DPLM family \cite{wang2024diffusion,wang2024dplm2,hsieh2025dplm2_1,wang2026towards} has sought to bridge this gap, noting that the strict left-to-right factorization of AR models may hinder the modeling of complex global residue interactions.
Despite these advances, existing protein foundation models still largely rely on predefined conditioning schemes, making them suboptimal for flexible instruction-driven reasoning. 
Consequently, protein understanding and sequence design are often addressed using separate models or task-specific fine-tuning pipelines. This limitation directly motivates \method, which unifies biological reasoning and conditional sequence design within a single foundation model equipped with open-ended natural language interfaces.

\paragraph{Large Language Models (LLMs) for Biology.}
Recent efforts have sought to overcome the rigid interfaces of specialized protein foundation models by integrating biological modalities into general-purpose LLMs.
One line of work focuses on \emph{modality adaptation}, where early approaches primarily projected textual annotations as conditioning input into specialized PLMs for controllable generation or retrieval-augmented prediction \cite{liu2023text,liu2025protein,Dai2024.08.01.606258}, leaving text encoder the major information bottleneck.
More recent works instead move toward the other way round, projecting proteins, nucleic acids, and other biological signals into the token space of established LLMs to enable biological understanding and instruction following \citep{NatureLM2025,fallahpour2025bioreason}.
Another direction studies \emph{agentic scientific systems}, where frontier LLMs are augmented with external biological harnesses.
Systems such as Claude for Life Sciences \citep{anthropic2025claude_for_life_sciences}, GPT-Rosalind \cite{openai2026gptrosalind} and other agentic frameworks \cite{huang2025biomni,jin2025stella} have shown great potential in multi-step biological problem-solving via tool calling.
While such orchestration is immensely powerful, its potential is naturally bounded by the native biological capabilities of the foundation model. 
Motivated by this, our work targets the gap between general LLMs and specialized PLMs.

%% file: sections/06conclusion.tex
\section{Conclusion}
\label{sec:conclusion}

In this work, we argue that biology needs models with \emph{native protein capability}: the ability to process proteins directly from sequence, rather than relying mainly on external tools or retrieved knowledge. Our analysis shows that strong general language ability does not automatically translate into strong protein competence: under a shared evaluation protocol, frontier LLMs still lag behind protein-specialized models and bioinformatics tools on sequence-grounded biological understanding and design.

To address this gap, we present \method, a unified protein--text foundation model that treats proteins as a native modality within a diffusion language modeling framework. By casting both protein understanding and sequence design as instruction-following tasks over a shared token space, \method supports diverse tasks instead of separate task-specific pipelines, while encouraging protein knowledge to be internalized and transferred across tasks.
We also introduce \bench, a benchmark for evaluating native protein capability under realistic generalization settings. \bench adopts time-aware and homology-aware protocols across both understanding and design tasks, enabling direct comparison among baselines under a common evaluation setup.
On \bench, \method generally outperforms frontier LLMs and shows strong competitiveness with protein-specialized models and bioinformatics tools. 

Taken together, these results suggest that unified protein--text modeling is a promising route toward more general scientific foundation models, reducing the fragmentation of task-specific workflows. We hope that releasing \method and \bench will support future research on scientific discovery.

%% file: sections/contributions.tex
\newpage
\section*{Contributions}

\textbf{Project Lead}

Lihao Wang$^{1,2}$

\textbf{Model}

Keyue Qiu$^{1,2,3}$, Yixin Wu$^{1,2,5}$, Zihan Zhou$^{1,7}$, Changze Lv$^{1,5}$, Lihao Wang$^{1,2}$



\textbf{Evaluation}

Yawen Ouyang$^{1,2}$, Jixiang Yu$^{6}$, Dongyu Xue$^{1,2}$





\textbf{Other Contributors}

 Yuxuan Song$^{2,3}$, Xinbo Zhang$^{1,2}$, Hao Wang$^{2,3}$, Jiangtao Feng$^{1,2,3}$, Zhiqiang Gao$^{1}$, Lijun Wu$^{1}$, Xiaoqing Zheng$^{5}$, Ka-Chun Wong$^{6}$, Lei Bai$^{1}$, Ya-Qin Zhang$^{3}$, Wei-Ying Ma$^{3,6}$, Dahua Lin$^{1}$, Bowen Zhou$^{1,4}$

\textbf{Co-first Authors}

Keyue Qiu$^{1,2,3}$, Yixin Wu$^{1,2,5}$

\textbf{Correspondence}

Hao Zhou$^{1,2,3}$

\subsection*{Affiliation}

$^1$Shanghai Artificial Intelligence Laboratory

$^2$Generative Symbolic Intelligence Lab (GenSI), Tsinghua University

$^3$Institute for AI Industry Research (AIR), Tsinghua University

$^4$Tsinghua University

$^5$Fudan University

$^6$City University of Hong Kong

$^7$Chinese University of Hong Kong, Shenzhen

\section*{Acknowledgments}

We thank DeepLink Team for their infrastructure support. 
This work is supported by Shanghai Artificial Intelligence Laboratory. 

%% file: sections/appendix.tex

\section{Methodology Notation}

In \Cref{tab:notation}, we summarize and provide corresponding descriptions for the notations used when introducing our block-wise diffusion large language model backbone \method~dLLM in \Cref{sec:dllm_architecture}.

\begin{table}[h!]
\centering
\begin{tabular}{ll}
\toprule
\textbf{Symbol} & \textbf{Description} \\
\midrule
$\mathcal{V}$ & Shared discrete vocabulary \\
$K$ & Vocabulary size (excluding \texttt{[MASK]}) \\
$\mathbf{m}$ & Absorbing \texttt{[MASK]} token vector ($e_{K+1}$) \\
$x_0$ & Clean token (one-hot vector) \\
$x_{i,0}$ & Clean token at position $i$ \\
$\mathbf{x}_0$ & Clean sequence \\
$\tilde{x}_t$ & Corrupted token at time step $t$ \\
$\tilde{x}_{i,t}$ & Corrupted token at position $i$ and time $t$ \\
$\tilde{\mathbf{x}}_t$ & Corrupted sequence at time step $t$ \\
$\alpha_t$ & Monotonically decreasing noise schedule at step $t$ \\
$Q_t$ & Transition matrix for the forward process \\
$\mathbf{x}$ & Interleaved full sequence \\
$L$ & Total number of partitioned contiguous blocks \\
$D$ & Number of tokens per block \\
$B_k$ & The $k$-th clean block \\
$B_{<k}$ & Preceding clean or fully denoised blocks $(B_1, \ldots, B_{k-1})$ \\
$\widetilde{B}_{k,t}$ & Corrupted $k$-th block at noise level $t$ \\
$M_{k,t}$ & Set of masked positions in block $k$ at time $t$ \\
$S$ & Interleaved training sequence of clean and corrupted blocks \\
$A_{uv}$ & Binary attention mask between positions $u$ and $v$ \\
\bottomrule
\end{tabular}
\caption{Summary of notations used in the dLLM architecture.}
\label{tab:notation}
\end{table}

\vspace{-6mm}

\begin{figure}[b!]
    \centering
    \includegraphics[width=0.96\linewidth]{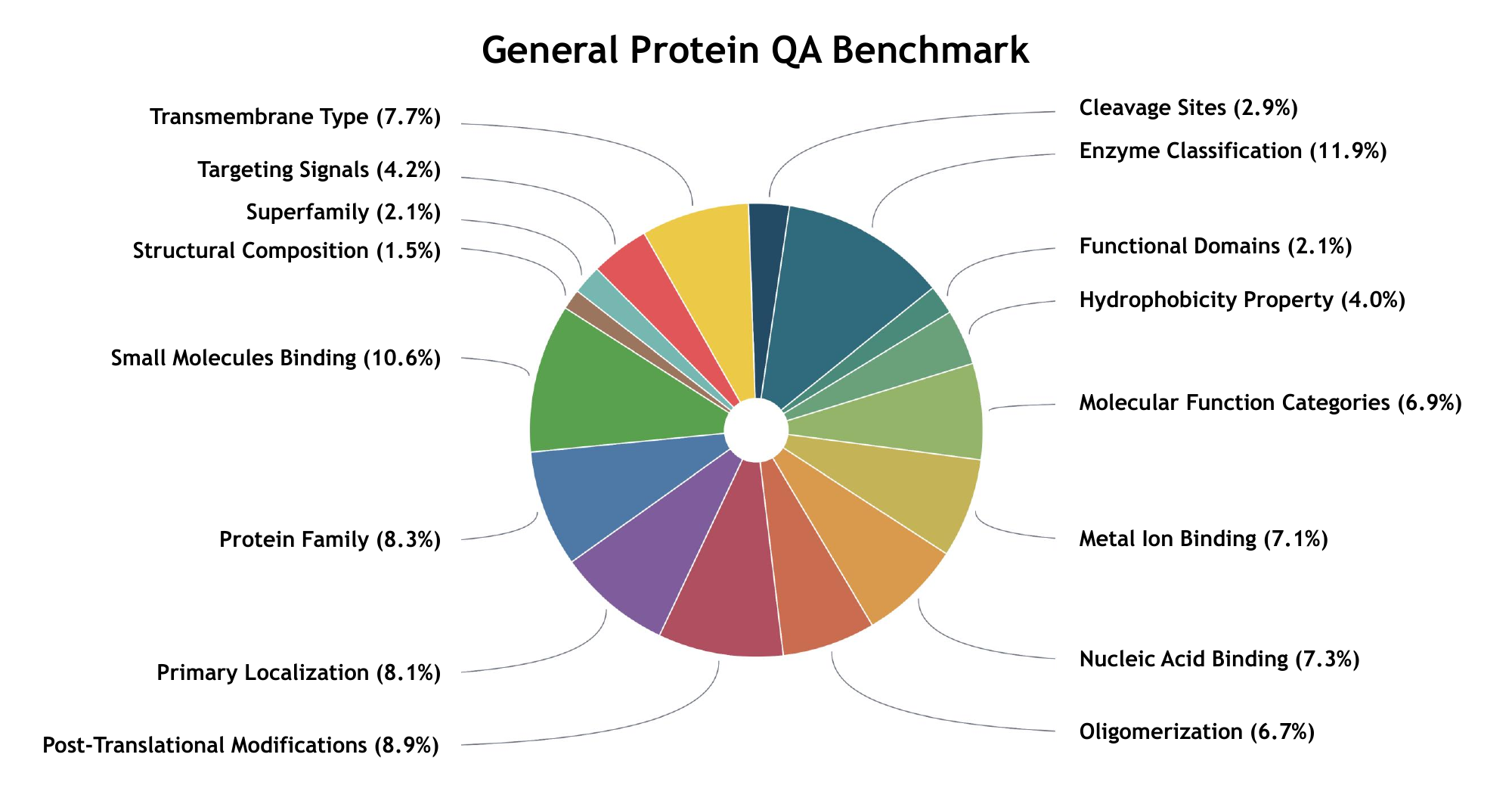}
    \vspace{-3mm}
    \caption{Task distribution of the General Protein QA track of \bench.}
    \vspace{-3mm}
    \label{fig:general_qa_benchmark}
\end{figure}

\newpage
\section{Data Instances}

\subsection{General Protein QA Tasks}
\label{app:subsec:general_protein_qa_data}

We present representative questions for each of the 16 protein understanding categories in \Cref{tab:general_qa_inst}, which are further paraphrased before being included for evaluation in \bench. These tasks are categorized into 5 core dimensions of protein knowledge: Function, Interaction and Binding, Location and Modification, Physicochemical Property, and Structure. This multi-dimensional taxonomy ensures a holistic evaluation of the model's capacity to internalize complex biological priors and generalize across the vast landscape of protein research.

\begin{table}[h!]
\small
\renewcommand{\arraystretch}{1.05}
\centering
\begin{tabularx}{\textwidth}{p{3.2cm} p{2.2cm} X}
\toprule
\textbf{Task Category} & \textbf{Aspect} & \textbf{Representative Question} \\ \hline

\addlinespace[0.5em] 
        \textbf{Enzyme} \newline \textbf{Classification}               & Function                      & Assign the top-level enzyme class (oxidoreductase, transferase, hydrolase, lyase, isomerase, ligase, or translocase) or declare non-enzymatic if no enzymatic function is indicated. \\ 
    \addlinespace[0.4em] 
        \textbf{Functional Domains}                  & Function                      & What functional domains are identified in this protein? \\
    \addlinespace[0.4em] 
        \textbf{Molecular Function} \newline \textbf{Categories}       & Function    & Infer the single most specific molecular function category, e.g. enzyme, transporter, ion channel, receptor, nucleic acid-binding, regulator, structural molecule, motor protein, or other. \\ 
    \addlinespace[0.4em] 
        \textbf{Protein Family}                      & Function                      & Which protein family does this sequence belong to? \\
    \addlinespace[0.4em] 
        \textbf{Superfamily}                         & Function                      & Assign the most plausible evolutionary superfamily membership. \\ 
    \addlinespace[0.4em] 
        \textbf{Metal Ion Binding}                   & Interaction and Binding       & Is the amino acid sequence likely to coordinate metal ions? If yes, name the most probable metal type when determinable. (e.g. Zinc ions, Calcium ions, Iron ions, No metal binding) \\ 
    \addlinespace[0.4em] 
        \textbf{Nucleic Acid} \newline \textbf{Binding}                & Interaction and Binding       & Does this protein bind nucleic acids and what type? (e.g. DNA-binding, RNA-binding, both DNA and RNA binding, no nucleic acid binding) \\ 
    \addlinespace[0.4em] 
        \textbf{Oligomerization}                     & Interaction and Binding       & Infer the most probable oligomerization state (e.g., monomer, homodimer, heterodimer, higher-order oligomer) for the protein. \\ 
    \addlinespace[0.4em] 
        \textbf{Small Molecules} \newline \textbf{Binding}             & Interaction and Binding       & Judge whether this protein binds to small-molecule, and if yes, name the single most probable cofactor type. \\ 
    \addlinespace[0.4em] 
        \textbf{Cleavage Sites}                      & Location and Modification     & What types of cleavage motifs are present? (e.g. signal peptide cleavage, propeptide cleavage, protease sites, no cleavage sites) \\ 
    \addlinespace[0.4em] 
        \textbf{Post-Translational} \newline \textbf{Modifications}    & Location and Modification     & What types of post-translational modifications are likely? (e.g. phosphorylation, glycosylation, ubiquitination, methylation, acetylation, multiple modifications, no PTMs predicted) \\ 
    \addlinespace[0.4em] 
        \textbf{Primary Localization}                & Location and Modification     & Where is this protein primarily localized in the cell? (e.g. cytosol, nucleus, membrane, secreted, mitochondrion, other) \\
    \addlinespace[0.4em] 
        \textbf{Targeting Signals}                   & Location and Modification     & What targeting signals are present in this protein? (e.g. signal peptide, mitochondrial targeting sequence, nuclear localization signal, no targeting signals predicted) \\
    \addlinespace[0.4em] 
        \textbf{Hydrophobicity} \newline \textbf{Property}             & Physicochemical Property      & What is the hydrophobic character of this protein? (e.g. highly hydrophobic, amphipathic, mostly hydrophilic, mixed regions) \\ 
    \addlinespace[0.4em] 
        \textbf{Structural} \newline \textbf{Composition}              & Structure                     & What broad structural fold class does this protein belong to? (e.g. all-alpha, all-beta, alpha/beta, alpha+beta, membrane protein, other) \\
    \addlinespace[0.4em] 
        \textbf{Transmembrane} \newline \textbf{Type}                  & Structure                     & Is this a transmembrane protein and what type? (e.g. single-pass, multi-pass, no transmembrane regions) \\
\addlinespace[0.5em] 

\bottomrule
\end{tabularx}
\caption{Task categories and representative questions for General Protein QA.}
\label{tab:general_qa_inst}
\end{table}

\subsection{Instruction-Tuning Data}

We provide data cases for the tasks of General Protein QA, Functional De Novo Design, EC Prediction, and CATH Prediction in \Cref{tab:understanding_example,tab:denovo_example,tab:ec_example,tab:cath_example} to better depict the objectives of each task and their differences.

\begin{table}[b!]
    \centering
    \small
    \renewcommand{\arraystretch}{1.05}
\begin{tabularx}{\textwidth}{l X}
    \toprule
    \textbf{Field} & \textbf{Content} \\
    \midrule
    \textbf{Instruction} & Determine the principal subcellular compartment in which this protein is found. The protein is {\ttfamily \seqsplit{MNIFNQLKQDIIVASRQLYNNQEIANTATIEIPKDSFNGDLSSNIAMIIAAKESIAPREVALKFKEVLITLPYIASIEIAGPGFINFTIKADSWQASIKDILQHEEKFFEIDIDKSRNINIEYVSANPTGPMHIGHARGAVYGDVLARILQKVSYSVTKEYYVNDAGSQINDLVSTVLLRYKEALGEQITIPAGLYPGEYLIPLGQILAKEYGNKLLTMNYAERFKIIKSFAVEKMLDLNRKDLADLGIKHDIFFSEQSLHDKGEIEETVKLLERMGLIYEGTLPAPKGKIHEEWDNRVQKLFKSTKYGDSQDRPIEKADGSWSYFASDLAYAKDKIERGANHLIYVLGADHSGYVKRIEAIVKALGKEQVKVDVKICQLVNFVENGVPVKMSKRLGSFASVQDVNHEVGKDIIRFMMLTRQNDKPLDFDLVKVKEQSRENPIFYVQYAHVRTISILSKAKELMPESYNNFESGKYDLSLLSSEEEIEIIKLLASWTKTLEASAKYFEPHRIAFYLINLASKFHSMWNFGKENSDYRFVIESNKELTLARLALASAIQKVIASGLEVIGVEPMNKM}}. \\
    \midrule
    \textbf{Answer} & Cytoplasm (cytosol) \\
    \midrule
    \textbf{Rationale} & The protein sequence lacks an N‑terminal signal peptide or transmembrane helices, indicating it is not secreted or membrane‑bound. The protein belongs to the class‑I aminoacyl‑tRNA synthetase family, enzymes that function in the cytosol to charge tRNAs during protein synthesis. The conserved “HIGH” motif ({\ttfamily \seqsplit{PTGPMHIGH}}) typical of class‑I tRNA synthetases is present, confirming its identity as an Arg‑tRNA ligase. Such synthetases are typically cytoplasmic in bacteria and eukaryotes, and the absence of organelle‑targeting signals (e.g., mitochondrial presequences) further supports a cytosolic location. Therefore, the most plausible primary localization for this ArgRS is the cytoplasm (cytosol).\\
    \bottomrule
\end{tabularx}
\caption{Data instance for General Protein QA.}
\label{tab:understanding_example}
\end{table}

\begin{table}[b!]
    \centering
    \small
    \renewcommand{\arraystretch}{1.05}
\begin{tabularx}{\textwidth}{l X}
    \toprule
    \textbf{Field} & \textbf{Content} \\
    \midrule
    \textbf{Instruction} & Create a biologically plausible complete protein sequence that would exhibit the characteristic properties and functions of the KOW protein domain. Consider the typical structural features, length, and functional characteristics associated. \\
    \midrule
    \textbf{Answer} & The {\ttfamily \seqsplit{KNDTVVVLSGDDKGKQGAVLELIPAKKAAIV}} is a conserved KOW domain characterized by a $\beta$‑barrel fold and clusters of basic residues for RNA binding; it could be constructed into the full-length {\ttfamily \seqsplit{MGKIRKNDTVVVLSGDDKGKQGAVLELIPAKKAAIVKGVNIKTKHRKPSNKNTSGEIITFEAPILLSKLALVAKKATKDKPAIPTRVGFKIENKKKIRIAKKTGKAI}}. \\
    \midrule
    \textbf{Rationale} & The design requirements specify a protein that contains a KOW domain within a compact basic RNA-binding protein of roughly 70--100 aa. KOW domains are common in ribosomal proteins. \newline
    The {\ttfamily \seqsplit{KNDTVVVLSGDDKGKQGAVLELIPAKKAAIV}} segment satisfies these requirements. Its key features include:
    \begin{itemize}
        \item A conserved G-rich loop ({\ttfamily{G}}) that forms part of the $\beta$-barrel core.
        \item Basic Lysine ({\ttfamily{K}}) residues distributed to create a positive RNA-binding surface.
        \item A pattern of alternating hydrophobic ({\ttfamily{V L}}) and polar residues that stabilizes the barrel.
    \end{itemize}
    To build the complete protein, combine: \newline
    1. An N-terminal basic tail rich in ({\ttfamily{K R}}) that may enhance RNA association. \newline
    2. The central KOW-like {\ttfamily \seqsplit{KNDTVVVLSGDDKGKQGAVLELIPAKKAAIV}} segment providing the $\beta$-barrel structure. \newline
    3. A C-terminal helix with conserved ({\ttfamily{A L}})-rich residues to complete the ribosomal protein fold. \newline
    Connecting these elements yields a plausible full protein sequence, denoted as {\ttfamily \seqsplit{MGKIRKNDTVVVLSGDDKGKQGAVLELIPAKKAAIVKGVNIKTKHRKPSNKNTSGEIITFEAPILLSKLALVAKKATKDKPAIPTRVGFKIENKKKIRIAKKTGKAI}}, which meets the length and functional criteria for a KOW‑containing ribosomal protein. \\
    \bottomrule
\end{tabularx}
\caption{Data instance for Functional De Novo Design.}
\label{tab:denovo_example}
\end{table}

For protein understanding tasks, question-answer pairs are extracted based on relevant protein context from curated Swiss-Prot annotations within UniProtKB~\cite{10.1093/nar/gkac1052} and CARE~\cite{yang2024care}. These questions are concatenated with their corresponding protein sequences to form comprehensive training instructions. To bolster the reasoning capabilities of \method and facilitate the acquisition of generalizable biochemical principles, we incorporate step-by-step rationales derived from and verified against the protein records. 
For the design task, instructions are formulated based on functional constraints from InterPro~\cite{blum2025interpro}, with answers and rationales synthesized according to the motif fragments and specific protein characteristics under strict fidelity validation. 
Amino acids are tokenized independently from natural language, establishing protein sequences as a native modality within the model.

\begin{table}[b!]
    \centering
    \small
    \renewcommand{\arraystretch}{1.05}
\begin{tabularx}{\textwidth}{l X}
    \toprule
    \textbf{Field} & \textbf{Content} \\
    \midrule
    \textbf{Instruction} & Determine the most appropriate four‑level EC number for the protein whose amino‑acid sequence is provided. The protein is {\ttfamily \seqsplit{MTTSDLPAFWTVIPAAGVGSRMRADRPKQYLDLAGRTVIERTLDCFLEHPMLRGLVVCLAEDDPYWPGLDCAASRHVQRAAGGVERADSVLSGLLRLLELGARADDWVLVHDAARPNLTRGDLDRLLEELAEDPVGGLLAVPARDTLKRSDRDGRVSETIDRSVVWLAYTPQMFRLGALHRALADALVAGVAITDEASAMEWAGYAPKLVEGRADNLKITTPEDLLRLQRSFPH}}. \\
    \midrule
    \textbf{Answer} & 2.7.7.60 \\
    \midrule
    \textbf{Rationale} & The protein is 234 residues long, which matches the typical size (220–240 aa) of enzymes belonging to the IspD family (2‑C‑methyl‑D‑erythritol 4‑phosphate cytidylyltransferase). IspD enzymes are cytidylyltransferases that catalyze the transfer of a nucleotidyl group to a phosphorylated substrate, a reaction classified under transferases that transfer phosphorus‑containing groups. Sequence analysis reveals motifs characteristic of nucleotidyltransferases, such as the conserved {\ttfamily{HDAA}} segment and a downstream {\ttfamily{HRAL}} motif, both of which are commonly found in IspD proteins and are required for binding CTP and the phosphorylated substrate. The overall fold predicted from the length and conserved residues is consistent with the cytidylyltransferase (CTP‑binding) family, placing the enzyme in EC class 2 (transferases), subclass 7 (transferring phosphorus‑containing groups), sub‑subclass 7 (nucleotidyltransferases). The specific activity described for IspD—conversion of 2‑C‑methyl‑D‑erythritol 4‑phosphate to its CDP‑activated form—is uniquely assigned the serial number 60 in this sub‑subclass. Therefore, the most likely EC number for this sequence is 2.7.7.60. \\
    \bottomrule
\end{tabularx}
\caption{Data instance for EC Prediction.}
\label{tab:ec_example}
\end{table}

\begin{table}[b!]
    \centering
    \small
    \renewcommand{\arraystretch}{1.05}
\begin{tabularx}{\textwidth}{l X}
    \toprule
    \textbf{Field} & \textbf{Content} \\
    \midrule
    \textbf{Instruction} & Determine the most probable CATH hierarchical classification (x.x.x.x) for the provided protein sequence. The protein is {\ttfamily \seqsplit{MLNTELLTTGVKGLDELLGGGVAKGVILQVYGPFATGKTTFAMQVGLLNEGKVAYVDTEGGFSPERLAQMAESRNLDVEKALEKFVIFEPMDLNEQRQVIARLKNIVNEKFSLVVVDSFTAHYRAEGSREYGELSKQLQVLQWIARRKNVAVIVVNQVYYDSNSGILKPIAEHTLGYKTKDILRFERLRVGVRIAVLERHRFRPEGGMVYFKITDKGLEDVKNED}}. \\
    \midrule
    \textbf{Answer} & 3.40.50.300 \\
    \midrule
    \textbf{Rationale} & The sequence is 225 residues long, a size typical for small nucleotide‑binding enzymes. A close inspection shows the classic Walker A P‑loop motif: {\ttfamily{GPFATGKTT}}, where the conserved {\ttfamily{GxxxxGKT}} pattern is present. Further downstream a Walker B acidic motif {\ttfamily{DE}} can be identified, confirming the presence of a P‑loop NTP‑binding domain. Such domains are built from alternating $\beta$‑strands and $\alpha$‑helices, placing them in CATH class 3 (Alpha‑Beta). The overall arrangement matches the 3‑Layer (aba) Sandwich architecture, corresponding to CATH level 2 code 3.40. The connectivity of the $\beta$‑sheet and surrounding helices follows the Rossmann‑like topology, giving level 3 code 3.40.50. Finally, the combination of the Walker motifs and the Rossmann topology is characteristic of the P‑loop containing nucleotide triphosphate hydrolases superfamily, which is assigned CATH level 4 code 300. Therefore, the most likely CATH number for this protein is 3.40.50.300. \\
    \bottomrule
\end{tabularx}
\caption{Data instance for CATH Prediction.}
\label{tab:cath_example}
\end{table}

\section{Experimental Setup}
\subsection{Inference Setting}\label{app:subsec:baseline}
\paragraph{Bio Tools}
Bio-tool baselines are employed as retrieval methods, and evaluated on the enzyme commission (EC) prediction and CATH structural fold classification tasks. 
The reference proteins are indexed into a database, with each query sequence searched against that database, and the annotation of the top-1 hit is transferred as the prediction.

\textbf{BLAST-DIAMOND} is used as a representative sequence-based retrieval tool, following the CARE~\cite{yang2024care} preprocessing utilities. The database is built from the UniProt enzyme reference FASTA and thus precludes non-enzyme candidates. 

\textbf{Foldseek}~\cite{van2024fast} is used for structure-based search against the prepared AlphaFoldDB-V4 database \citep{10.1093/nar/gkad1011}.

\paragraph{Bio Models}
Bio-model baselines are evaluated on EC, CATH, and when supported by the model interface, InterPro-guided functional design. 

\textbf{ESM2}\footnote{\url{https://github.com/facebookresearch/esm}}~\cite{lin2023evolutionary} is a large-scale transformer-based protein language model trained with a masked language modeling objective to internalize evolutionary patterns from hundreds of millions of diverse sequences. In this study, we employ the \texttt{esm2-t33-650M-UR50D} checkpoint to extract high-dimensional protein representations, which are subsequently fed into a task-specific trainable head for downstream predictions.

\textbf{ESM3}\footnote{\url{https://github.com/evolutionaryscale/esm}}~\cite{hayes2025simulating} is a large-scale generative masked protein language model that jointly reasons over sequence, structure, and function tracks represented as discrete tokens, and fills masked positions through iterative sampling. It can be used for both understanding and design tasks, and is run with the \texttt{ESM3-sm-open-v1} checkpoint. For protein design, ESM3 is function-annotation-conditioned rather than instruction-conditioned. The primary protocol keeps only InterPro IDs supported by the ESM3 tokenizer and builds the precisely spanned \texttt{FunctionAnnotation} constraints. ESM3 supports a larger subset of InterPro labels compared to CFP-Gen, covering 826/870 samples.

\textbf{ProTrek}\footnote{\url{https://github.com/westlake-repl/ProTrek}}~\cite{su2024protrek} is a tri-modal protein language model that enables joint contrastive learning across protein sequence, structure, and function tracks. By integrating a pre-trained ESM encoder for amino acid sequences, a BERT-based text encoder for functional descriptions, and a structural encoder for Foldseek-derived 3Di tokens, ProTrek maps multiple modalities into a shared representation space. In this work, we focus on its sequence and structure modalities by concatenating the representations using the \texttt{ProTrek-650M-UniRef50} checkpoint.

\textbf{SaProt}\footnote{\url{https://github.com/westlake-repl/SaProt}}~\cite{su2023saprot} is a structure-aware protein language model that utilizes a novel ``structure-aware'' (SA) vocabulary, which integrates residue types with structural information. By encoding 3D structures into discrete tokens using Foldseek, SaProt represents proteins as sequences of SA tokens, combining primary sequence data with tertiary geometric conformations. 
We utilize the \texttt{SaProt-650M-AF2} checkpoint to extract representations.

\textbf{ProteinDT}\footnote{\url{https://github.com/chao1224/ProteinDT}}~\citep{liu2023text} is a text-guided protein design framework with three stages: \texttt{ProteinCLAP} aligns protein and text representations through contrastive learning, a facilitator maps text representations toward the protein latent space, and a decoder generates protein sequences from the resulting representation. The model is text-conditioned, but its upstream text encoder is trained on SwissProtCLAP-style UniProt free text rather than imperative design instructions. Therefore, the primary protocol uses prompts constructed from matched UniProt comment text. The prompt builder concatenates available UniProt XML text comments, including function, catalytic activity, pathway, subunit, similarity, and cofactor descriptions. The selected decoder variant is T5, a Transformer-based autoregressive decoder, using the released \texttt{ProtBERT-BFD/SciBERT} text-protein checkpoint and T5 decoder checkpoint. Inference uses default setting and one explicit seed per design.

\textbf{Pinal}\footnote{\url{https://github.com/westlake-repl/Denovo-Pinal}}~\cite{Dai2024.08.01.606258} is a natural-language-to-protein design baseline that first converts a text description into structural tokens with T2struc and then uses SaProt-T to generate amino-acid sequences conditioned on the predicted structure and text. Pinal accepts the compact InterPro-derived caption text description, namely a semicolon-separated list of \texttt{<InterPro-Name> <InterPro-Type>} phrases. The loaded model components are the released \texttt{T2struc-1.2B} and \texttt{SaProt-T 760M} weights.

\textbf{Chroma}\footnote{\url{https://github.com/generatebio/chroma}}~\citep{ingraham2023illuminating} is a programmable generative model for proteins that uses diffusion modeling with equivariant graph neural networks and conditional random fields to sample all-atom structures; its sequence and side-chain design networks are jointly trained with the structure model. Chroma is not conditioned on the raw benchmark instruction in the primary protocol. Instead, captions are passed to \texttt{ProCapConditioner}, which provides natural-language conditioning as a differentiable conditioner during sampling. The selected replicated variant uses ProCap-aligned UniProt-style captions; direct InterPro name/type captions and instruction-only captions are used only for ablations. We follow the default setting and use the released Chroma backbone, design, and ProCap checkpoints with the \texttt{GPT-Neo-125M} caption model.

\textbf{CFP-Gen}\footnote{\url{https://github.com/yinjunbo/cfpgen}}~\citep{yin2025cfpgen} is a combinatorial functional protein generation baseline built on the diffusion language model DPLM \citep{wang2024diffusion}. It introduces annotation-guided feature modulation for composable functional labels and residue-controlled functional encoding for residue-wise control, supporting GO, InterPro, EC, sequence-motif, and backbone constraints. CFP-Gen is class-conditional rather than natural-language-conditioned. The current protocol conditions only on directly supported InterPro identifiers from the denovo benchmark; no InterPro identifier is rescued through Gene Ontology mappings. Supported InterPro IDs are mapped to CFP-Gen class IDs, and targets without a supported InterPro ID are dropped. For each seed, the generated config uses the \texttt{CFP-Gen 650M} checkpoint following default configuration. 
CFP-Gen fails to support most of the newly incorporated InterPro labels, and thus only 319 samples out of 870 have been evaluated.

\textbf{ProDVa}\footnote{\url{https://github.com/sornkL/ProDVa}}~\citep{liu2025protein} is a dynamic-vocabulary augmented protein design baseline that combines a text encoder for functional descriptions, a protein language model for sequence generation, and a fragment encoder/retriever that injects protein fragments as a task-specific dynamic vocabulary. The functional-design protocol uses CAMEO-style keyword-list prompts and ProDVa's retrieval-backed dynamic vocabulary. The active configuration uses the ProDVa-CAMEO checkpoint, CAMEO protein-fragment phrases, and PubMedBERT embeddings. Inference follows default configuration.

\begin{table}[t]
    \centering
    \small
    \renewcommand{\arraystretch}{1.05}
\begin{tabular}{p{0.12\linewidth}p{0.26\linewidth}p{0.54\linewidth}}
\toprule
\textbf{Model} & \textbf{Conditioning Type} & \textbf{Representative Prompt} \\
\midrule
\addlinespace[0.5em]
\textbf{CFP-Gen} & Class-label conditioning (supported labels only) & \texttt{IPR008567}, \texttt{IPR013785}. \\
\addlinespace[0.4em]
\textbf{ESM3} & Class-label conditioning as function annotation (supported labels only) & \texttt{FunctionAnnotation(label=IPR008567, start=S1, end=E1), FunctionAnnotation(label=IPR013785, start=S2, end=E2)}. \\
\addlinespace[0.4em]
\textbf{Chroma} & Text prompt from UniProt functional comments via \texttt{ProCapConditioner} caption & Catalyzes the condensation of 3,5-dioxohexanoate and acetyl-CoA, forming acetoacetate and acetoacetyl-CoA. May be involved in fatty acid biosynthesis rescue via triacetic acid lactone. Belongs to the BKACE family. \\
\addlinespace[0.4em]
\textbf{ProteinDT} & Text prompt from UniProt comments & Catalyzes the condensation of 3,5-dioxohexanoate and acetyl-CoA, forming acetoacetate and acetoacetyl-CoA. May be involved in fatty acid biosynthesis rescue via triacetic acid lactone. 3,5-dioxohexanoate + acetyl-CoA = acetoacetyl-CoA + acetoacetate. Belongs to the BKACE family. \\
\addlinespace[0.4em]
\textbf{Pinal} & Text prompt from InterPro labels & Beta-keto acid cleavage enzyme family; Aldolase-type TIM barrel homologous superfamily; Beta-keto acid cleavage enzyme family \\
\addlinespace[0.4em]
\textbf{ProDVa} & Text prompt constructed from InterPro labels & Generate a protein sequence for a novel protein that integrates the following function keywords: Beta-keto acid cleavage enzyme, Aldolase-type TIM barrel. The designed protein sequence is \\
\addlinespace[0.5em]
\bottomrule
\end{tabular}
\caption{Representative conditioning inputs built for the bio-model functional-design baselines. The examples correspond to the same denovo target with InterPro annotations \texttt{IPR008567} (BKACE) and \texttt{IPR013785} (Aldolase TIM).}
\label{tab:bio_model_conditioning_examples}
\end{table}

\paragraph{Frontier LLMs}
Frontier LLMs are evaluated across all the protein understanding tasks and the functional-design task. We use default temperature setting, with \texttt{top\_p} also omitted unless explicitly required by the API or reasoning mode (e.g., \texttt{top\_p=0.95} for Claude). We set \texttt{max\_tokens=8192} by default, and relax it to \texttt{32768} only when the task output may exceed the default limit. For functional design, the model is additionally required to return exactly one uppercase amino-acid sequence with length at most 1024 residues.

\paragraph{\method}
    For protein understanding tasks, \method~dLLM uses \texttt{temperature=0.7}, \texttt{top\_p=0.9} and \texttt{max\_tokens=2048}, while \method~AR uses \texttt{temperature=0.7}, \texttt{top\_p=0.5} and \texttt{max\_tokens=4096}. For protein design, both models use \texttt{temperature=0.7} and \texttt{top\_p=0.6}, which we find more suitable for generation over the much smaller amino-acid vocabulary (20 amino acid tokens). Across tasks, we find that temperatures around 0.6--0.8 work well in practice; top-\textit{p} around 0.6 is generally preferable for protein design, while 0.6--0.9 works fine for protein understanding. Note that the meaning of temperature is similar between AR and dLLM, as both use it to control the sharpness of the sampling distribution, but for top-\textit{p} it differs: for AR, it follows standard nucleus sampling based on the cumulative next-token probability mass, whereas for dLLM, it controls truncation of marginal probabilities across parallel tokens at each iterative denoising step, making it less susceptible to compounding exposure bias.

\subsection{Evaluation Metrics}\label{app:subsec:eval_setting}

\paragraph{Functional Design}
In addition to the accuracy adopted for protein understanding, we detail the metrics calculation for functional design as follows, where each method is evaluated in terms of sequence repetition, foldability, functional recovery, and novelty.


For sequence repetition, we report \texttt{Repeat} \citep{kuang2025pdfbench}, which measures the fraction of residues contained in tandemly repeated regions, together with \texttt{Rep2} and \texttt{Rep5} following Rep-N~\citep{welleck2019neural}. For n-gram repetition, we compute
\[
\mathrm{Rep}_n(x) = 1 - \frac{|\mathrm{unique}(G_n(x))|}{|G_n(x)|},
\]
where \(G_n(x)\) is the multiset of all overlapping \(n\)-grams in sequence \(x\). We report \texttt{Rep2} and \texttt{Rep5} by setting \(n=2\) and \(n=5\), respectively, and averaging over generated sequences.

For structural plausibility, we use \texttt{ESMFold-v1} \citep{lin2023evolutionary} as the folding model to calculate pLDDT.

For sequence-level similarity metrics, we employ MMseqs2 \citep{steinegger2017mmseqs2} for sequence identity against all UniProt sequences up until \textbf{December 31, 2025} \citep{10.1093/nar/gkae1010}, following \citet{kuang2025pdfbench}. Novelty is calculated as the nearest-neighbor sequence novelty:
\[
\mathrm{Novelty}(x) = 1 - \max_y \mathrm{SeqID}(x,y),
\]
Note that for the Swiss-Prot test sequence (Natural), we intentionally remove the identical matches as the sequence database being searched against contains the reference sequence itself.


For functional recovery, we first run \texttt{InterProScan-5.75-106.0}\footnote{https://ftp.ebi.ac.uk/pub/software/unix/iprscan/5/5.75-106.0/} \citep{10.1093/bioinformatics/btu031} over generated sequences.
Then, we compute the recovery rate between the generated sequence $s$ and reference sequence $s_{\text{ref}}$ as
\[
\mathrm{Recovery\ Rate} =
\begin{cases}
\dfrac{
\left|\operatorname{InterProScan}(s)\cap \operatorname{InterProScan}(s_{\text{ref}})\right|
}{
\left|\operatorname{InterProScan}(s_{\text{ref}})\right|
},
& \text{if } \operatorname{InterProScan}(s_{\text{ref}})\neq \varnothing, \\[8pt]
\mathrm{N/A},
& \text{if } \operatorname{InterProScan}(s_{\text{ref}})= \varnothing.
\end{cases}
\]

\section{Extended Experimental Results}
\label{app:subsec:homology_identity}

For the \textbf{Functional De Novo Design} task, we provide additional focused results on the intersection of samples shared by all baselines.
For the hierarchical classification tasks of \textbf{EC Prediction} and \textbf{CATH Prediction}, we provide evaluation results on proteins corresponding to a greater sequence identity split than 30\% with \bench. This indicates cases where the proteins evaluated share sequences with higher similarity to those readily trained for each model, thus resulting in better overall performance.

\paragraph{Functional De Novo Design} 
As described in Appendix \Cref{app:subsec:baseline}, protein-specialized language models impose more restrictive input requirements, being inherently limited by its training functional keywords and cannot generalize to unseen data, therefore some of them cannot be evaluated on the full test set. To enable a better controlled apples-to-apples comparison, we report in \Cref{tab:design_performance_comparison_intersection} results on the 207-sample intersection subset commonly evaluated by all baselines. This subset is generally easier for protein-specialized models, since corresponding functional keywords are in-distribution under the evaluation settings of all included baselines. Accordingly, both model groups improve on this subset: Bio-models gain an average of 2.48 in pLDDT and 9.05\% in IPR, while LLMs improve by 2.05 in pLDDT and 8.52\% in IPR.

The overall trend remains unchanged. Protein-specialized models may achieve stronger pLDDT, but are generally less effective at controlling functional keywords, suggesting limitations in projecting textual constraints into the protein space. On the other hand, as exemplified by \method~dLLM and Claude Opus 4.7, LLMs properly trained to capture the protein modality can enjoy the advantage of both high-fidelity protein sequence modeling, and high IPR, which is supported by their strong instruction-following capabilities.


\begin{table}[b!]
\centering
\small
\renewcommand{\arraystretch}{1.05}

\begin{tabular}{l >{\rmfamily}l 
SSS SS SS
} 
\toprule
    \multirow{2}{*}{\textbf{Category}}
    & \multirow{2}{*}{\textbf{Model}}
    & \multicolumn{3}{c}{\textbf{Repetition}}
    & \multicolumn{2}{c}{\textbf{Quality}}
    & \multicolumn{2}{c}{\textbf{Seq Distribution}} \\
\cmidrule(lr){3-5} \cmidrule(lr){6-7} \cmidrule(lr){8-9}
    & 
    & \textbf{Rep} & \textbf{Rep2} & \textbf{Rep5} & \textbf{pLDDT} & \textbf{IPR (\%)} & \textbf{Novelty} & \textbf{Unique (\%)} \\
\midrule

\multirow{1}{*}{\textbf{Reference}}
& Natural & 1.78 & 44.68 & 0.14 & 86.06 & 100.00 & 4.19 & 100.00 \\
\midrule

\addlinespace[0.3em]
\multirow{6}{*}{\textbf{Bio-Model}}
& CFPGen          & 10.98 & 57.86 & 10.79 & \ul{78.48} & 34.00 & 50.05 & 100.00 \\
& Chroma          &  2.41 & 56.49 &  0.47 & 58.69 &  0.08 & 59.21 & 100.00 \\
& ESM3            & 23.79 & 71.36 & 17.85 & 64.84 & 31.51 & 67.68 & 100.00 \\
& Pinal           & 11.49 & 52.28 &  3.01 & 75.40 & \ul{59.57} & 48.49 & 100.00 \\
& ProDVa          &  7.41 & 30.92 & 11.62 & 71.97 & 32.32 & 17.66 &  86.43 \\
& ProteinDT       &  2.57 & 51.60 &  0.34 & 39.11 & 21.55 & 50.40 & 100.00 \\
\addlinespace[0.1em]
\midrule
\addlinespace[0.5em]
\multirow{9}{*}{\textbf{LLM}}
& Qwen3.5-27B     & 13.28 & 86.45 & 74.35 & 29.97 &  0.12 & 79.25 & 100.00 \\
& GLM-5.1         &  1.21 & 48.08 &  1.18 & 33.98 &  7.95 & 51.36 & 100.00 \\
& DeepSeek‐V4‐Pro &  4.96 & 53.66 &  3.98 & 45.78 & 37.15 & 43.35 & 100.00 \\
& GPT‐5.5         &  2.52 & 55.84 &  2.03 & 40.22 & 35.88 & 34.92 & 100.00 \\
& Gemini 3.1 Pro  &  2.30 & 46.02 &  1.75 & 47.96 & 31.90 & 40.80 &  99.52 \\
& Claude Opus 4.6 &  0.98 & 47.27 &  0.34 & 43.47 & 35.42 & 38.71 &  95.17 \\
& Claude Opus 4.7 &  1.79 & 43.83 &  0.05 & 81.14 & \textbf{77.83} & 35.54 & 100.00 \\

\addlinespace[0.1em]
& \cc \textbf{\method~AR}   & \cc 1.63 & \cc 45.26 & \cc 0.26 & \cc 75.60 & \cc 55.45 & \cc 13.67 & \cc 51.21 \\
& \cc \textbf{\method~dLLM} & \cc 3.14 & \cc 45.34 & \cc 0.37 & \cc \textbf{81.62} & \cc 72.25 & \cc 30.49 & \cc 96.62 \\
\addlinespace[0.1em]

\bottomrule
\addlinespace[0.2em]
\multicolumn{9}{l}{\scriptsize* Results are recomputed on the 207 out of 870 reference sequences shared across all listed baselines.} \\
\end{tabular}
\caption{Intersection-only comparison of various models on generation quality, novelty, and uniqueness metrics.}
\label{tab:design_performance_comparison_intersection}
\end{table}

\paragraph{EC Prediction} 
Results in high-homology regimes ($\ge 30\%$ sequence identity) are detailed in \Cref{tab:performance_ec_full}. In these scenarios, protein-specialized models
and bioinformatics tools typically define the performance ceiling due to their direct exploitation of evolutionary conservation. Notably, \method outperforms frontier LLMs across all high-homology regimes, with \method~AR being nearly on par with specialized models like ESM2 and ESM3. This indicates that while most general LLMs fail to utilize high-homology templates effectively, \method successfully bridges the gap between generative language modeling and classical bioinformatics.

\paragraph{CATH Prediction} 
Evaluation results in \Cref{tab:performance_cath_full} reveals a sharp contrast between generative LLMs and structure-aware protein models and tools. All six baselines compared achieved near-perfect scores across all identity tiers, as they are specifically optimized for structural fold recognition. However, within the LLM cohort, \method demonstrates a dominant lead, greatly improving over the strongest frontier Claude Opus 4.7. This suggests that \method's training regime allows it to recognize structural hierarchies from sequences alone far more effectively than standard LLMs, establishing it as a highly reliable generative proxy for structural classification when 3D coordinates are unavailable.

\begin{table}[t]
\centering
\small
\renewcommand{\arraystretch}{1.05}

\begin{tabular}{l l >{\rmfamily}l SSSS} 
\toprule
    \textbf{Identity}
    & \textbf{Method Type}
    & \textbf{Model}
    & \textbf{Level 1}
    & \textbf{Level 2}
    & \textbf{Level 3}
    & \textbf{Level 4} \\
\midrule

\multirow{15}{*}{\textbf{30\%-50\%}}
& \multirow{2}{*}{\textbf{Bio-Tool}}
& BLAST             & 84.26 & 77.78 & 75.00 & 44.44 \\
& & Foldseek        & \ul{90.74} & \ul{85.19} & \ul{83.33} & \ul{49.07} \\

\cmidrule{2-7} 

& \multirow{4}{*}{\textbf{Bio-Model}}
& ESM2              & \ul{85.19} & \ul{79.63} & \ul{78.70} & \ul{44.44} \\
& & ESM3            & \ul{85.19} & 75.00 & 72.22 & 41.67 \\
& & ProTrek         & 81.48 & 73.15 & 71.30 & 38.89 \\
& & SaProt          & 83.33 & 74.07 & 71.30 & 41.67 \\

\cmidrule{2-7}

& \multirow{9}{*}{\textbf{LLM}}
& Qwen3.5-27B       & 6.48 & 1.85 & 1.85 & 0.00 \\
& & GLM-5.1         & 44.44 & 28.70 & 22.22 & 8.33 \\
& & DeepSeek-V4-Pro & 61.11 & 49.07 & 46.30 & 26.85 \\
& & GPT-5.5         & 42.59 & 28.70 & 23.15 & 16.67 \\
& & Gemini 3.1 Pro  & 61.11 & 49.07 & 44.44 & 21.30 \\
& & Claude Opus 4.6 & 54.63 & 37.04 & 33.33 & 18.52 \\
& & Claude Opus 4.7 & 79.63 & 70.37 & 65.74 & 32.41 \\

\addlinespace[0.1em]
& & \cc \textbf{\method~AR}   & \cc 80.56 & \cc \textbf{74.07} & \cc \textbf{73.15} & \cc \textbf{40.74} \\
& & \cc \textbf{\method~dLLM} 
& \cc \textbf{82.41} & \cc 73.15 & \cc 70.37 & \cc 38.89  \\
\addlinespace[0.1em]

\midrule

\multirow{15}{*}{\textbf{50\%-70\%}}
& \multirow{2}{*}{\textbf{Bio-Tool}}
& BLAST             & \ul{96.84} & \ul{89.47} & \ul{88.42} & \ul{77.89} \\
& & Foldseek        & \ul{96.84} & 88.42 & 87.37 & 76.84 \\

\cmidrule{2-7} 

& \multirow{4}{*}{\textbf{Bio-Model}}
& ESM2              & \ul{96.84} & \ul{91.58} & \ul{90.53} & \ul{78.95} \\
& & ESM3            & 95.79 & 87.37 & 87.37 & 74.74 \\
& & ProTrek         & 93.68 & 87.37 & 85.26 & 75.79 \\
& & SaProt          & \ul{96.84} & 89.47 & 88.42 & 76.84 \\

\cmidrule{2-7}

& \multirow{9}{*}{\textbf{LLM}}
& Qwen3.5-27B       & 14.74 & 4.21 & 4.21 & 0.00 \\
& & GLM-5.1         & 37.89 & 21.05 & 17.89 & 9.47 \\
& & DeepSeek-V4-Pro & 70.53 & 60.00 & 55.79 & 32.63 \\
& & GPT-5.5         & 57.89 & 43.16 & 42.11 & 24.21 \\
& & Gemini 3.1 Pro  & 68.42 & 55.79 & 52.63 & 30.53 \\
& & Claude Opus 4.6 & 49.47 & 31.58 & 28.42 & 11.58 \\
& & Claude Opus 4.7 & 92.63 & 83.16 & 83.16 & 60.00 \\

\addlinespace[0.1em]
& & \cc \textbf{\method~AR}   & \cc \textbf{95.79} & \cc 87.37 & \cc \textbf{87.37} & \cc \textbf{72.63} \\
& & \cc \textbf{\method~dLLM}
& \cc \textbf{95.79} & \cc \textbf{89.47} & \cc \textbf{87.37} & \cc \textbf{72.63}   \\
\addlinespace[0.1em]

\midrule

\multirow{15}{*}{\textbf{70\%-100\%}}
& \multirow{2}{*}{\textbf{Bio-Tool}}
& BLAST             & \ul{98.75} & \ul{97.50} & \ul{96.25} & \ul{88.75} \\
& & Foldseek        & 97.50 & 96.25 & \ul{96.25} & 87.50 \\

\cmidrule{2-7} 

& \multirow{4}{*}{\textbf{Bio-Model}}
& ESM2              & \ul{98.75} & \ul{96.25} & \ul{95.00} & \ul{85.00} \\
& & ESM3            & 93.75 & 90.00 & 87.50 & 77.50 \\
& & ProTrek         & 95.00 & 92.50 & 90.00 & 78.75 \\
& & SaProt          & 93.75 & 90.00 & 87.50 & 76.25 \\

\cmidrule{2-7}

& \multirow{9}{*}{\textbf{LLM}}
& Qwen3.5-27B       & 16.25 & 6.25 & 5.00 & 0.00 \\
& & GLM-5.1         & 52.50 & 35.00 & 28.75 & 13.75 \\
& & DeepSeek-V4-Pro & 83.75 & 63.75 & 63.75 & 41.25 \\
& & GPT-5.5         & 65.00 & 52.50 & 50.00 & 30.00 \\
& & Gemini 3.1 Pro  & 70.00 & 56.25 & 55.00 & 33.75 \\
& & Claude Opus 4.6 & 67.50 & 52.50 & 47.50 & 21.25 \\
& & Claude Opus 4.7 & 92.50 & 83.75 & 81.25 & 58.75 \\

\addlinespace[0.1em]
& & \cc \textbf{\method~AR}   & \cc 93.75 & \cc \textbf{90.00} & \cc \textbf{87.50} & \cc \textbf{76.25} \\
& & \cc \textbf{\method~dLLM} 
& \cc \textbf{95.00} & \cc 88.75 & \cc \textbf{87.50} & \cc 75.00 \\

\addlinespace[0.1em]

\bottomrule
\end{tabular}
\caption{Level-wise EC prediction accuracy on proteins with $\ge30\%$ sequence identity.}
\label{tab:performance_ec_full}
\end{table}

\begin{table}[t]
\centering
\small
\renewcommand{\arraystretch}{1.05}

\begin{tabular}{l l >{\rmfamily}l SSSSS} 
\toprule
    \textbf{Identity}
    & \textbf{Method Type}
    & \textbf{Model}
    & \textbf{Level 1}
    & \textbf{Level 2}
    & \textbf{Level 3}
    & \textbf{Level 4} \\
\midrule

\multirow{15}{*}{\textbf{30\%-50\%}}
& \multirow{2}{*}{\textbf{Bio-Tool}}
& BLAST             & 85.87 & 85.87 & 85.87 & 85.87 \\
& & Foldseek        & \ul{100.00} & \ul{100.00} & \ul{100.00} & \ul{100.00} \\

\cmidrule{2-7} 

& \multirow{4}{*}{\textbf{Bio-Model}}
& ESM2              & 98.91 & \ul{98.91} & \ul{97.83} & \ul{97.83} \\
& & ESM3            & 95.65 & 93.48 & 90.22 & 90.22 \\
& & ProTrek         & \ul{100.00} & 97.83 & 96.74 & 96.74 \\
& & SaProt          & 96.74 & 96.74 & 95.62 & 95.62 \\

\cmidrule{2-7}

& \multirow{9}{*}{\textbf{LLM}}
& Qwen3.5-27B       & 19.57 & 13.04 & 3.26 & 1.09 \\
& & GLM-5.1         & 56.52 & 28.26 & 17.39 & 7.61 \\
& & DeepSeek-V4-Pro & 65.22 & 40.22 & 32.61 & 23.91 \\
& & GPT-5.5         & 68.48 & 41.30 & 30.43 & 19.57 \\
& & Gemini 3.1 Pro  & 76.09 & 54.35 & 38.04 & 23.91 \\
& & Claude Opus 4.6 & 71.74 & 43.48 & 29.35 & 23.91 \\
& & Claude Opus 4.7 & 89.13 & 80.43 & 73.91 & 66.30 \\

\addlinespace[0.1em]
& & \cc \textbf{\method~AR}   & \cc 93.48 & \cc \textbf{83.70} & \cc \textbf{76.09} & \cc \textbf{72.83} \\
& & \cc \textbf{\method~dLLM} 
& \cc \textbf{97.83} & \cc 80.43 & \cc \textbf{76.09} & \cc 68.48 \\
\addlinespace[0.1em]

\midrule

\multirow{15}{*}{\textbf{50\%-70\%}}
& \multirow{2}{*}{\textbf{Bio-Tool}}
& BLAST             & \ul{100.00} & \ul{100.00} & \ul{100.00} & \ul{100.00} \\
& & Foldseek        & \ul{100.00} & \ul{100.00} & \ul{100.00} & \ul{100.00} \\

\cmidrule{2-7} 

& \multirow{4}{*}{\textbf{Bio-Model}}
& ESM2              & 98.39 & 98.39 & 98.39 & 98.39 \\
& & ESM3            & \ul{100.00} & \ul{100.00} & \ul{100.00} & \ul{100.00} \\
& & ProTrek         & \ul{100.00} & \ul{100.00} & \ul{100.00} & \ul{100.00} \\
& & SaProt          & \ul{100.00} & \ul{100.00} & \ul{100.00} & \ul{100.00} \\

\cmidrule{2-7}

& \multirow{9}{*}{\textbf{LLM}}
& Qwen3.5-27B       & 33.87 & 11.29 & 0.00 & 0.00 \\
& & GLM-5.1         & 66.13 & 35.48 & 16.13 & 6.45 \\
& & DeepSeek-V4-Pro & 83.87 & 58.06 & 53.23 & 45.16 \\
& & GPT-5.5         & 75.81 & 43.55 & 35.48 & 22.58 \\
& & Gemini 3.1 Pro  & 83.87 & 64.52 & 48.39 & 37.10 \\
& & Claude Opus 4.6 & 75.81 & 51.61 & 43.55 & 25.81 \\
& & Claude Opus 4.7 & 91.94 & 87.10 & 83.87 & 82.26 \\

\addlinespace[0.1em]
& & \cc \textbf{\method~AR}   & \cc \textbf{98.39} & \cc \textbf{98.39} & \cc \textbf{96.77} & \cc \textbf{96.77} \\
& & \cc \textbf{\method~dLLM} 
& \cc 91.94 & \cc 85.48 & \cc 83.87 & \cc 80.65 \\
\addlinespace[0.1em]

\midrule

\multirow{15}{*}{\textbf{70\%-100\%}}
& \multirow{2}{*}{\textbf{Bio-Tool}}
& BLAST             & \ul{100.00} & \ul{100.00} & \ul{100.00} & \ul{100.00} \\
& & Foldseek        & \ul{100.00} & \ul{100.00} & \ul{100.00} & \ul{100.00} \\

\cmidrule{2-7} 

& \multirow{4}{*}{\textbf{Bio-Model}}
& ESM2              & \ul{100.00} & \ul{100.00} & \ul{100.00} & \ul{100.00} \\
& & ESM3            & \ul{100.00} & \ul{100.00} & \ul{100.00} & \ul{100.00} \\
& & ProTrek         & \ul{100.00} & \ul{100.00} & \ul{100.00} & \ul{100.00} \\
& & SaProt          & \ul{100.00} & \ul{100.00} & \ul{100.00} & \ul{100.00} \\

\cmidrule{2-7}

& \multirow{9}{*}{\textbf{LLM}}
& Qwen3.5-27B       & 32.43 & 12.16 & 2.70 & 0.00 \\
& & GLM-5.1         & 56.76 & 22.97 & 13.51 & 6.76 \\
& & DeepSeek-V4-Pro & 71.62 & 55.41 & 48.65 & 31.08 \\
& & GPT-5.5         & 63.51 & 45.95 & 33.78 & 20.27 \\
& & Gemini 3.1 Pro  & 75.68 & 48.65 & 35.14 & 27.03 \\
& & Claude Opus 4.6 & 68.92 & 47.30 & 32.43 & 20.27 \\
& & Claude Opus 4.7 & 85.14 & 72.97 & 71.62 & 64.86 \\

\addlinespace[0.1em]
& & \cc \textbf{\method~AR}   & \cc 97.30 & \cc \textbf{95.95} & \cc \textbf{95.95} & \cc \textbf{93.24} \\
& & \cc \textbf{\method~dLLM} 
& \cc \textbf{98.65} & \cc 94.59 & \cc 94.59 & \cc 90.54 \\
\addlinespace[0.1em]

\bottomrule
\end{tabular}
\caption{Level-wise CATH prediction accuracy on proteins with $\ge30\%$ sequence identity.}
\label{tab:performance_cath_full}
\end{table}